\documentclass[pre]{revtex4}
\usepackage{epsfig}
\usepackage{bm}
\begin{document}
\title{Phase ordering with a global conservation law: Ostwald ripening and coalescence}
\author{Massimo Conti$^1$, Baruch Meerson$^2$, Avner Peleg$^2$ and
 Pavel V. Sasorov$^3$}
\affiliation{$^1$ Dipartimento di Matematica e Fisica, Universit\`{a} di Camerino,\\
and Istituto Nazionale di Fisica della Materia, 62032,
Camerino, Italy\\
$^2$The Racah Institute of Physics, The Hebrew University of Jerusalem, Jerusalem
91904, Israel \\
$^3$Institute for Theoretical and Experimental Physics, Moscow, 117259, Russia }
\begin{abstract}
Globally conserved phase ordering dynamics is investigated in systems with short range
correlations at $t=0$. A Ginzburg-Landau equation with a global conservation law  is employed as
the phase field model. The conditions are found under which the sharp-interface limit of this
equation is reducible to the area-preserving motion by curvature. Numerical simulations show that,
for both critical and off-critical quench, the equal time pair correlation function exhibits
dynamic scaling, and the characteristic coarsening length obeys $l(t)\sim t^{1/2}$. For the
critical quench, our results are in excellent agreement with earlier results. For off-critical
quench (Ostwald ripening) we investigate the dynamics of the size distribution function of the
minority phase domains. The simulations show that, at large times, this distribution function has
a self-similar form with growth exponent $1/2$. The scaled distribution, however, strongly differs
from the classical Wagner distribution. We attribute this difference to coalescence of domains. A
new theory of Ostwald ripening is developed that takes into account binary coalescence events. The
theoretical scaled distribution function agrees well with that obtained in the simulations.
\end{abstract}

\maketitle

\section{Introduction}
\label{intro}

Phase ordering is emergence of order from disorder through domain growth and coarsening. The
standard setting when phase ordering occurs is a temperature quench from a high-temperature
disordered phase into a two-phase or a multiphase region. Phase ordering has been the subject of
extensive research during the last two decades \cite{Bray94}. An important simplifying assumption
in phase ordering theory is dynamic scale invariance. According to this assumption, the coarsening
system possesses, at late times, a single relevant dynamic length scale $l(t)$ (the characteristic
domain size) which grows with time as $l(t)\sim t^{\alpha}$ \cite{Bray94}. It is by now well
established that in systems with short range correlations $\alpha=1/2$ for non-conserved (model A)
dynamics, while $\alpha=1/3$ for locally conserved (model B) dynamics .

There is, however, an important additional coarsening mechanism:
\textit{globally conserved} phase ordering
\cite{Schimansky,Rubinstein,Majumdar,MS,Rutenberg,PCM}. Globally
conserved dynamics can be thought of as model A dynamics
constrained by global conservation of the order parameter: for
example, Ising model with fixed magnetization. This global
conservation law is maintained by an external field (for example,
a magnetic field) which depends on time but is uniform in space.
The globally-conserved phase ordering is accessible in experiment.
Consider the sublimation/deposition dynamics of a solid and its
vapor in a small closed vessel kept at a constant temperature
below the melting point. As the acoustic time scale in the gas
phase is short compared to the coarsening time, the gas pressure
(and, consequently, density) remain uniform in space, changing
only in time. This character of mass transport in the vapor phase
makes the coarsening dynamics conserved globally rather than
locally. An important characteristics of globally-conserved
dynamics is interface-controlled kinetics, in contrast to the
bulk-diffusion-controlled kinetics typical for locally-conserved
systems. Interface-controlled kinetics was investigated in the
context of growth of small platinum particles supported on alumina
substrates in an oxidizing environment \cite{Wynblatt}. There are
additional examples of cluster growth on surfaces
\cite{Zinke-Allmang}, where it was found possible to single out
the interface-controlled kinetics \cite{Williams}. There is also
strong evidence in favor of globally conserved
interface-controlled transport during the coarsening of clusters
in granular powders driven by a low-frequency electric field
\cite{Aranson1,AMSV}.

Part of the theoretical importance of globally conserved phase ordering lies in the fact that it
enables an access to off-critical quenches in the simpler model A dynamics (with global
conservation). Thus, it allows one to determine which characteristics of the system depend on the
volume (or area) fraction $\varepsilon$ and which do not.

Dynamic renormalization group arguments show that global conservation should not change the growth
law \cite{Bray91}. This early result was confirmed by particle simulations with short range
correlations in the initial conditions: for critical ($\varepsilon=1/2$)
\cite{Rutenberg,simulations} and for off-critical ($\varepsilon<1/2$) \cite{Majumdar} quench.
Recent phase field simulations of systems with long-range (power-law) correlations in the initial
conditions have also shown dynamic scale invariance with the same growth exponent $\alpha=1/2$
\cite{PCM}. Therefore, $\alpha=1/2$ independently of $\varepsilon$. On the other hand, the
autocorrelation function \cite{Majumdar,Lee} and persistence exponent \cite{Lee} were found to be
$\varepsilon$-dependent.

Globally conserved dynamics are related to a wide range of multiphase coarsening systems. Sire and
Majumdar \cite{Majumdar} showed that in the large-$q$ limit the dynamics of the $q$-state Potts
model are equivalent to the dynamics of the globally conserved model with an area fraction
$\varepsilon=1/q$. The large-$q$ limit of the Potts model is of practical importance as it
describes correctly some of the dynamic characteristics of dry soap froths \cite{Glazier} and of
the coarsening of polycrystalline materials \cite{Grest}.

In the limit of a vanishing volume fraction of the minority phase
the late stage of coarsening is describable by the mean field
theories of Ostwald ripening. Lifshitz and Slyozov \cite{LS}
developed such a theory for the bulk-diffusion-controlled (or
locally conserved) dynamics. They showed that the size
distribution function of the minority domains approaches, at large
times, a self-similar form. Correspondingly, the average size of
the minority phase domains grows with time like $t^{1/3}$.
Following the seminal work by Lifshitz and Slyozov \cite{LS},
Wagner developed a similar mean-field theory for the
interface-controlled (globally conserved) Ostwald ripening
\cite{Wagner}. The Wagner's theory yields a growth law $t^{1/2}$
for the average domain size, and a different (broader) shape of
the scaled distribution function. We shall refer to this scaled
distribution function as the Wagner distribution. More recently,
it was shown that interface-controlled Ostwald ripening appears in
the sharp-interface limit of scalar Ginzburg-Landau equations (and
its modifications) with a global conservation law
\cite{Schimansky,Rubinstein,Majumdar,MS,AMS}.

Although the simple theories of Lifshitz-Slyozov and Wagner were
developed more than 40 years ago, they are still very useful in
phase ordering theory. For example, Sire and Majumdar
\cite{Majumdar} employed the Wagner distribution to calculate the
equal-time pair correlation function. Lee and Rutenberg \cite{Lee}
used the two theories of Ostwald ripening for calculating the
autocorrelation exponent and persistence exponent for the locally-
and globally-conserved systems.

Many works were devoted to extensions of the Lifshitz-Slyozov theory to finite volume fractions.
Already Lifshitz and Slyozov \cite{LS} made an attempt to go beyond their simple model and account
for coalescence. Later it became clear that, in the locally-conserved systems, the dominant effect
unaccounted for by the simple theory is inter-domain correlations, rather than coalescence. At
small area fractions, the relative role of correlations is of order $\varepsilon^{1/2}$
\cite{Marder}, while the relative role of coalescence is of order $\varepsilon$. Therefore, an
account of coalescence without a proper account of correlations is an excess of accuracy.

The situation is quite different in globally-conserved systems, and this fact has not been
recognized until now. Correlations between neighboring domains are \textit{exponentially} small in
this case \cite{Rubinstein,MS}. Therefore, coalescence is expected to give the dominant correction
to the theory of Wagner \cite{Wagner}. We shall report numerical simulations that show a strong
effect of coalescence at moderate $\varepsilon$. Specifically, we find that, at large times, the
size distribution function of the minority domains has a self-similar form with the ``normal"
growth exponent $1/2$. The scaled distribution function, however, strongly differs from the Wagner
distribution. We attribute this difference to coalescence and develop a new theory of Ostwald
ripening that takes coalescence into account.

The outline of the rest of the paper is the following. In Section \ref{model} we briefly review the
phase field model of globally conserved phase ordering: a scalar Ginzburg-Landau equation with a
global conservation law. The sharp-interface asymptotic limit of this equation is introduced and
reduced, in 2D, to a simpler model of area-preserving motion by curvature. The criteria for the
validity of this reduction are obtained and presented in Appendix. The model of area-preserving
motion by curvature is used to obtain dynamic scaling laws for the characteristic coarsening
length and for the effective magnetic field. The results of the phase field simulations for the
critical quench ($\varepsilon=0.50$) and off-critical quench ($\varepsilon=0.25$) are presented and
analyzed in Section \ref{simu}. In Section \ref{sor} a new theory of Ostwald ripening is
developed. The theory leads to a non-linear integro-differential equation for the scaled
distribution function. The solution of this equation is in good agreement with the scaled size
distribution function found in the phase field simulations. In Section \ref{conclusions} we
summarize our results.

\section{Phase Field Model and Sharp-Interface Limit}
\label{model}

Globally conserved phase ordering dynamics are describable by a simple phase field model
\cite{Rubinstein,Majumdar,Rutenberg}. In this model the free energy functional has the
Ginzburg-Landau form:
\begin{equation}
F[u] = \int \left[\frac{\delta}{2\mu}(\nabla u)^2 + V(u) + H u
\right] d^{d} {\bf
  r} \,,
\label{newth1}
\end{equation}
and the dynamics follow a simple gradient descent:
\begin{equation}
\partial_t u =-\mu\,\frac{\delta F}{\delta u} =\delta\,\nabla^{2}u+\mu\,\\\left(
u-u^{3}- H(t) \right) \,.
 \label{newth2}
\end{equation}
Either no-flux or periodic boundary conditions can be used. In Eqs. (\ref{newth1}) and
(\ref{newth2}) $u({\mathbf{r}},t)$ is the coarse-grained order parameter field, $V(u) = (1/4)
(1-u^2)^2$ is a symmetric double-well potential, $\delta$ is the diffusion coefficient, $\mu$ is
the characteristic rate of relaxation of the field $u$ to its stable equilibrium values, and $d$ is
the dimension of space. The effective uniform magnetic field $H(t)$ changes in time so as to impose
the global conservation law:
\begin{equation}
\langle u({\mathbf{r}},t) \rangle = L^{-d}\int u({\mathbf{r}},t)\, d^{d}{\mathbf{r}}=\mbox{const}
 \,,
\label{newth3}
\end{equation}
where $L$ is the system size and the integration is carried out
over the entire system. Integrating both sides of Eq.
(\ref{newth2}) over the entire system and using Eq. (\ref{newth3})
and the boundary conditions we obtain:
\begin{equation}
H(t)= \langle u-u^3 \rangle=L^{-d}\int \left[u({\mathbf{r}},t)-u^{3}({\mathbf{r}},t)\right]
d^{d}{\mathbf{r}}
 \,.
\label{newth4}
\end{equation}
Therefore, Eq. (\ref{newth2}) takes the form:
\begin{equation}
\partial_t u = \delta\,\nabla^{2}u+\mu\,(u-u^{3})-\mu \langle u-u^3 \rangle \,,
\label{newth5}
\end{equation}
a globally-constrained Ginzburg-Landau equation  (GLE). From now on we shall concentrate on the 2D
case.

At late stages of the coarsening process, the system consists of domains of "phase 1" (where $u$
is close to $-1$) and "phase 2" (where $u$ is close to $1$) separated by domain walls. The domain
walls can be treated as sharp interfaces \cite{MS,Mikhailov}, as their characteristic width
$\lambda=(\delta/\mu)^{1/2}$ is much smaller than the characteristic domain size $l(t)$ which
grows with time. At this stage $H(t)$ is already small, $H(t)\ll 1$, and slowly varying in time.
The phase field in the phases 1 and 2 is almost uniform and rapidly adjusting to the value of
$H(t)$, so $u \simeq -1-H(t)/2$ and $u \simeq 1-H(t)/2$ in the phases 1 and 2, respectively. Under
these conditions, the so called``sharp-interface theory" holds. The normal velocity of the
interface $v_{n}$ is given by \cite{MS}:
\begin{equation}
v_{n}(s,t)=\delta\,\kappa(s,t)+(\delta\,\mu)^{1/2}\,gH(t)
 \,,
\label{newth6}
\end{equation}
where $s$ is a coordinate along the interface, $\kappa(s,t)$ is
the local curvature, and $g=-3/\sqrt{2}$. A positive $v_{n}$
corresponds to the interface moving towards phase 1, while a
positive  $\kappa$ corresponds to an interface which is convex
towards phase 2. The dynamics of $H(t)$ are described by
\cite{MS}:
\begin{equation}
\dot{H}(t)=\frac{4\Lambda(t)}{L^{2}}
\left[\delta\,\overline{\kappa(s,t)}+(\delta\mu)^{1/2}\,gH(t)\right]
 \,.
\label{newth7}
\end{equation}
Here $\overline{\kappa(s,t)}$ is the interface curvature averaged over the whole interface:
\begin{equation}
\overline{\kappa(s,t)}=\frac{1}{\Lambda(t)}\oint\kappa(s,t)\,ds
 \,,
\label{newth8}
\end{equation}
and $\Lambda(t) = \oint\,ds$ is the total perimeter of the interface. Equations (\ref{newth6}) and
(\ref{newth7}) provide a general sharp-interface formulation for the GLE with a global conservation
law.

Let us denote by $A(t)$ the total area of phase 2: $$
A(t)=\int_{u({\mathbf{r}},t)>0}d^{2}{\mathbf{r}}\,. $$ Equations
(\ref{newth6}) and (\ref{newth7}) can be used to calculate the
rate of change of $A(t)$:
\begin{equation}
\dot{A}(t) = \oint v_{n}(s,t)ds =
\Lambda(t)\left[\delta\,\overline{\kappa(s,t)}+(\delta\mu)^{1/2}\,gH(t)\right]
 \,.
 \label{newth9}
\end{equation}
Using Eqs. (\ref{newth7}) and (\ref{newth9}) we obtain: $\dot{H}=4\dot{A}/L^{2}$ which yields the
global conservation law:
\begin{equation}
A(t)-\frac{L^{2} H(t)}{4}=\mbox{const}
 \,.
\label{conservation}
\end{equation}
The second term in Eq. ~(\ref{conservation}) corresponds to the bulk order parameter $u$ being
biased by $H(t)$. One can use Eq.~(\ref{conservation}) instead of Eq. (\ref{newth7}) in the general
sharp-interface formulation of the problem.

In some important cases this formulation can be simplified further \cite{MS,PCM,PMVC}. When the two
terms on the right hand side of Eq. (\ref{newth9}) approximately balance each other,
\begin{equation}
H(t)\simeq -\frac{1}{g}\left(\frac{\delta}{\mu}\right)^{1/2}\,\overline{\kappa(s,t)}
 \,,
 \label{newth10}
\end{equation}
the area of each of the two phases remains constant. In this case Eq. (\ref {newth6}) takes the
form:
\begin{equation}
v_{n}(s,t)=\delta\,\left[\kappa(s,t)-\overline{\kappa(s,t)}\,\right]
 \,.
\label{newth11}
\end{equation}
Dynamics (\ref{newth11}) are known as area-preserving motion by curvature in 2D, and as
volume-preserving motion by mean curvature in 3D \cite{Rubinstein,MS,Gage}. Due to the presence of
the non-local term $\overline{\kappa}$ this model is different from the Allen-Cahn equation
\cite{Allen-Cahn} $v_{n}=\delta\,\kappa$, which represents the sharp-interface limit for
non-conserved (model A) dynamics \cite{Bray94}.

A simple example where the area-preserving dynamics \textit{cannot} be used is the dynamics of a
single circular domain of the minority phase in a ``sea" of the majority phase \cite{MS,AMS}.
Another example is the dynamics of a ``donut": a single domain of the minority phase with an
inclusion of a majority phase domain \cite{MM}. Therefore, the first question we need to address
concerns the general conditions under which the area-preserving dynamics, Eq. (\ref{newth11}),
represent an accurate approximation to the more general sharp-interface theory, Eqs.
(\ref{newth6}) and (\ref{newth7}). These conditions are derived in Appendix.

Now we employ the area-preserving model and do simple dynamic scaling analysis. (In the rest of
the paper we are using dimensionless variables  and put $\delta=\mu =1$.) For critical quench we
have $H(t)=0$ and $\overline{k(s,t)}=0$ because of symmetry between the two phases (we neglect
finite-size effects). Therefore, the globally conserved dynamics for critical quench are
\textit{identical} to the non-conserved (model A) dynamics. Using the Allen-Cahn equation
$v_{n}(s,t)=\kappa(s,t)$, one arrives at the well-known scaling law $l(t)\sim t^{1/2}$
\cite{Bray94}.

Turning to the off-critical quench, we notice that, under the scaling assumption, the interface
velocity can be estimated as $v_{n}\sim dl/dt$. Each of the two terms on the right hand side of
Eq. (\ref{newth11}) is of order $1/l(t)$. Equating and integrating, we again obtain $l(t)\sim
t^{1/2}$. Therefore, global conservation does not change the dynamic scaling for \textit{any} area
fraction. This result was previously obtained by dynamic renormalization group arguments applied to
Eq. (\ref{newth5}) (with a Gaussian white noise term) \cite{Bray91}, and by particle simulations
\cite{Majumdar}. For the off-critical dynamics of $H(t)$ we have: $|H(t)|=|<\kappa(s,t)>/g|\sim
1/l(t) \sim t^{-1/2}$.

Though the dynamic exponent is independent of the area fraction, other characteristics can depend
on it. In the following Section we report numerical simulations that address
area-fraction-dependent quantities.

\section{Numerical simulations}
\label{simu}

We performed extensive simulations by directly solving Eq.~(\ref{newth5}) with initial conditions
in the form of "white noise". The simulations were done for two different values of the area
fraction of the minority phase: $\varepsilon=0.50$, and $\varepsilon=0.25$, corresponding to a
critical and off-critical quench, respectively. In both cases the results were averaged over 10
different samples. Eq.~(\ref{newth5}) was discretized and solved on a $1024 \times 1024$ domain,
with mesh size $\Delta x=\Delta y=1$ and periodic boundary conditions. The coarsening process was
followed up to a time $t=3000$. An explicit Euler integration scheme was used to advance the
solution in time, and the Laplace operator was discretized by second order central differences. A
time step $\Delta t=0.1$ was required for numerical stability. The accuracy of the numerical
scheme was monitored by checking the (approximate) conservation law (\ref{conservation}) of the
general sharp-interface theory. It was found that this conservation law is obeyed with an accuracy
better than 0.02{\%} for $t>4$ and better than 0.008{\%} for $t>30$ in the critical quench case.
In the off-critical quench the approximate conservation law (\ref{conservation}) is obeyed with an
accuracy better than 1{\%} for $t>30$. To avoid any misunderstanding here and in the following we
notice that, in all cases, the \textit{integrated order parameter} [see Eq. (\ref{newth3})] is
conserved exactly by the numerical scheme.

\begin{figure}
\hspace{-0.1in} \center{\epsfysize=4.5in \epsffile{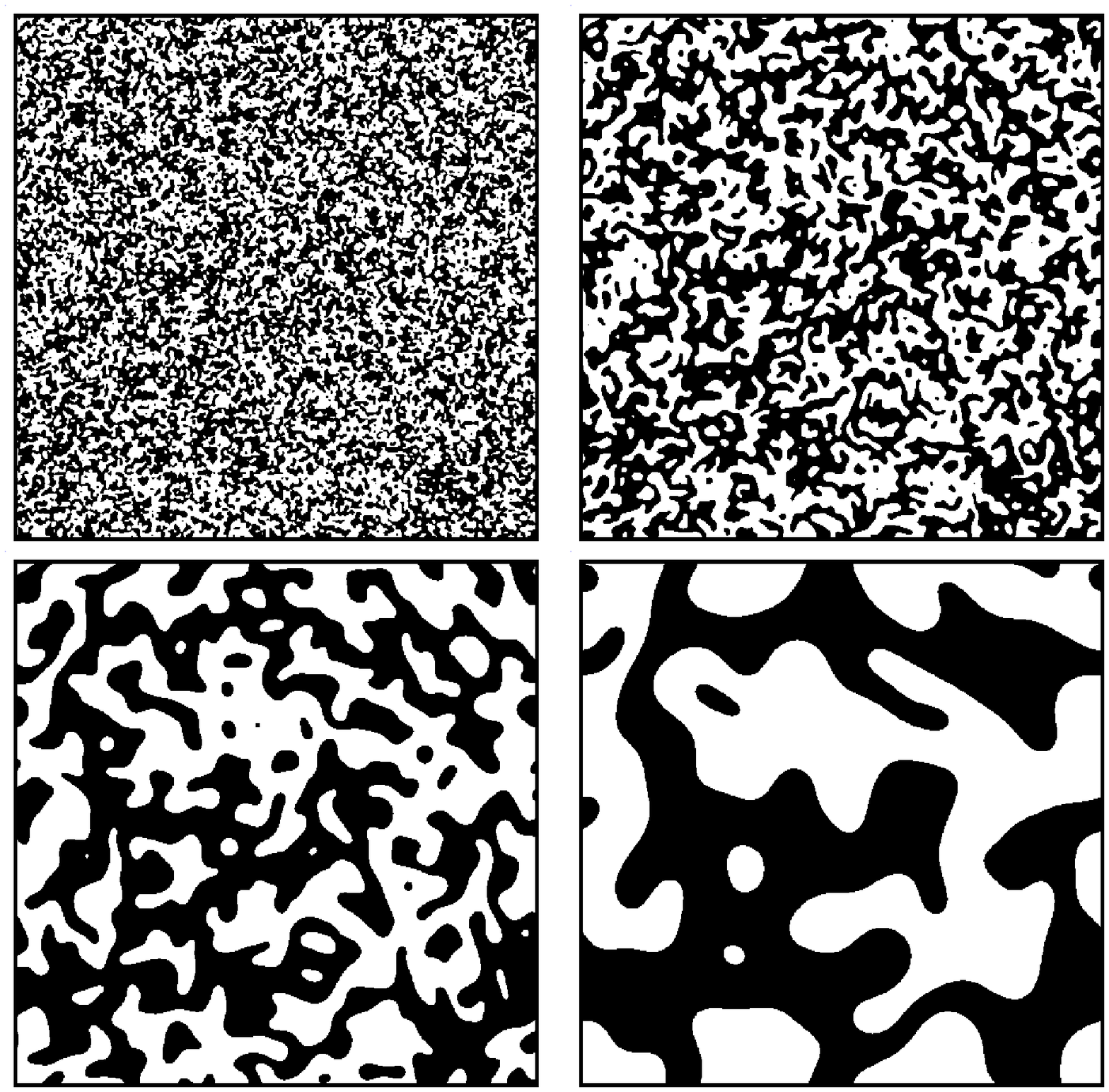}}\vspace{-1.3in}\caption{Snapshots
of globally conserved  coarsening for critical quench.  The upper row corresponds to $t=5.2$ (left)
and 32.3 (right), the lower row to $t=204.8$ (left) and 1305.5 (right). }\label{fig1}
\end{figure}

It is convenient to introduce an auxiliary density field $\rho({\mathbf r},t)=(1/2)[u({\mathbf
r},t)+1]$. The minority phase is identified as the locus where $\rho({\mathbf r},t)\geq 1/2$.
Typical snapshots of the coarsening process are shown in Fig. \ref{fig1} for the critical quench,
and in Fig. \ref{fig2} for the off-critical quench. For the critical quench the system consists of
interpenetrating domains of the two phases. For the off-critical quench the morphology is that of
Ostwald ripening \cite{Ostwald}: larger domains of the minority phase grow at the expense of
smaller ones. As the minority phase area fraction is not very small, binary (and even triple)
coalescence events are clearly seen in Fig. \ref{fig2}. Overall, the coarsening morphologies
resemble those observed for locally-conserved system: in numerical solutions of the Cahn-Hilliard
equation \cite{Rogers} and in particle simulations \cite{Fratzl}. An important difference is an
apparent absence of correlations between neighboring domains in Fig. \ref{fig2}.

To analyze the coarsening dynamics, the following quantities were sampled and averaged over the 10
initial conditions:
\begin{enumerate}
\item The area of phase 2.
\item The circularly averaged equal-time pair correlation
function:
\begin{equation}
C(r,t)= \frac{<\rho ({\mathbf{r'}},t)\rho ({\mathbf{r'}}+{\mathbf{r}},t)>- {<\rho
({\mathbf{r'}},t)>}^{2}} {<\rho ^{2}({\mathbf{r'}},t)>-{<\rho({\mathbf{r'}},t)>}^{2}}
 \,.
 \label{correlation}
\end{equation}
\item The characteristic coarsening length scale $l(t)$,
determined from the condition $C(l,t)=1/2$.
\item The effective magnetic field $H(t)$ computed from Eq. (\ref {newth4}).
\item The size distribution function of the minority phase domains
(for the off-critical quench).
\end{enumerate}

For critical quench we found that the area of the minority phase is constant with an accuracy
better than 0.03{\%} at all times. The situation is quite different for the off-critical quench.
Here there is a systematic trend in the area fraction of the minority phase. Still, with time this
quantity approaches a constant value. Deviations from this constant value become less than 3{\%}
for $t>100$. This approximate area conservation plays a crucial role in the theory of Ostwald
ripening (see Sec. \ref{sor}).

Figure \ref{fig3} shows, on a single graph, the scaling forms of the correlation function $C(x)$,
where $x=r/l(t)$, for the critical and off-critical quench. The  $l(t)$-dependence is presented in
Fig. \ref{fig4}. A comparison of the scaling forms $C(x)$ with those obtained in particle
simulations of globally conserved \cite{Rutenberg} and non-conserved \cite{Humayun} dynamics for
critical quench is also shown. The three curves for the critical quench almost coincide. For
off-critical quench, the $C(x)$ curve is slightly different from the curves for the critical
quench. A similar weak dependence of the scaled correlation function on the area fraction of the
minority phase was observed in locally conserved systems \cite{Rogers,Fratzl,Elder}.

Figure \ref{fig4} shows corrected power-law fits $l(t)=l_{0}+bt^{\alpha}$ which yield
$\alpha=0.50$, $l_{0}=0.5$, and $b=1.2$ for the critical quench, and $\alpha=0.51$, $l_{0}=1.3$,
and $b=0.9$ for the off-critical quench. A pure $t^{1/2}$ power-law line serves as a reference for
the expected late-time dynamic behavior. Therefore, $l(t)$ obeys the expected $t^{1/2}$ dynamical
scaling law, in agreement with the predictions of the dynamic renormalization group analysis
\cite{Bray91} and area-preserving sharp-interface theory. The difference in the values of the
amplitudes $b$ again
indicates a dependence on the minority phase area fraction. 

The time history of $1/|H(t)|$ for the off-critical quench is presented in Fig. \ref{fig5}. The
data is fitted by a corrected power-law: $1/|H(t)|=a+c\,t^{\alpha}$ with $a=7.4$, $c=2.3$, and
$\alpha=0.51$. Also shown is a $2.3\,t^{1/2}$ power law, serving as a reference to the expected
late-time dynamics. We conclude that $|H(t)|\sim t^{-1/2}$, as predicted by the sharp-interface
theory. The significance of the value of the amplitude $c$ will be discussed in Section \ref{sor}
in the context of our theory of Ostwald ripening with coalescence. One can distinguish in Fig.
\ref{fig5} small "fluctuations" of $1/|H(t)|$ around a smooth trend. (This is in contrast to the
$l(t)$-dependence where no fluctuations are observed.) To interpret these fluctuations we use Eqs.
(\ref{newth10}) and (\ref{eqtopo}) to obtain: $1/|H(t)|\sim\Lambda(t)/N_{2}(t)$. $\Lambda(t)$ is a
continuous function of $t$, whereas $N_{2}(t)$ behaves discontinuously at the time moments when
domains disappear due to shrinking and merging events. Thus $H(t)$ serves as a ``domain counter".

For critical quench, $H(t)$ exhibits very small irregular fluctuations around zero. The typical
values of $H(t)$ in this case are of the order of $10^{-5}$, and we interpret these fluctuations as
finite-size effects.

As we have shown, the scaled correlation function only weakly depends on the area fraction. A much
more sensitive diagnostics of the off-critical quench dynamics is provided by the size distribution
function of the minority phase domains. We found that, at late times, this function exhibits
dynamic scaling. Figure \ref{fig7} shows the scaled form $\Phi_{num}$ of the distribution function
obtained in the simulations with the GLE. The scaled variable on the horizontal axis of Fig.
\ref{fig7} is $\xi =R/t^{1/2}$, where the effective radius of each domain is defined as
$R=(A_d/\pi)^{1/2}$ and $A_d$ is the domain area. Function $\Phi_{num}$ was obtained, at each
moment of time, by multiplying the values of the distribution function, found in the simulations,
by $t^{3/2}$. The dynamic exponents $1/2$ and $3/2$ are the same as in the classical theory of
Wagner \cite{Wagner}.

Here is a more detailed account of our calculation of the scaled distribution function. We chose
for sampling 13 time moments in the interval $120<t<2900$. The domain statistics is obviously
better at earlier times of this interval, and it deteriorates at later times, as many domains
shrink and disappear. On the other hand, the dynamic exponent $1/2$ shows up, with a good
accuracy, only at relatively late times (see Figs. \ref{fig4} and \ref{fig5}). Therefore,  we had
to include the relatively late times in our sampling, which led to relatively big error bars in
Fig. \ref{fig7}.

The area fraction $\varepsilon=0.25$, used in our simulations, is moderately large. Therefore, one
could expect significant deviations of the scaled distribution function, found numerically, from
the Wagner distribution \cite{Wagner} corresponding to the same area fraction (that is, having the
same second moment). The Wagner distribution has the following form:
\begin{equation}
\Phi_W(\xi)=C\,\varepsilon\,\frac{\xi}{(\xi-\sqrt{2})^4}\, \exp\left(-\frac{2\sqrt{2}}{\sqrt{2}
-\xi}\right)  \label{wagner}
\end{equation}
for $\xi<\sqrt{2}$, and $\Phi_W (\xi) =0$ for $\xi\ge\sqrt{2}$. The normalization constant
$$
C=\frac{1}{\pi \left[(2 e^2)^{-1} + Ei (-2)\right]} \simeq 16.961\,,
$$
where $Ei \,(...)$ is the exponential integral function \cite{mathematica}.

The two distributions, $\Phi_{num} (\xi)$ and $\Phi_W (\xi)$, are shown in the same Fig.
~\ref{fig7}. One can see that the difference between them is enormous (in order to show the Wagner
distribution on the same graph with $\Phi_{num}$, we had to multiply it by a factor of 0.5).
Therefore, at moderate area fractions, the Wagner's theory is inapplicable.

It is instructive to compare the zero moments $M_0$ of the two distributions. The zero moment is
the amplitude of the scaling law for the number density of domains at large times:
$n(t)=M_0\,t^{-1}$. We obtained $M_0=4.72 \cdot 10^{-2}$ for $\Phi_{num}$ and $M_0 \simeq 1.43
\cdot 10^{-1}$ for $\Phi_W$. Therefore, for $\varepsilon=0.25$ the Wagner distribution
overestimates the number of domains at late times by a factor of $3$.  An additional difference is
the pronounced tail in $\Phi_{num}$ which extends much further than the edge of the compact
support of the Wagner distribution. Coalescence provides a natural explanation to these two facts:
coalescence events reduce the total number of domains and produce domains of progressively larger
size. We shall see in the next Section that an account of coalescence leads to a good quantitative
agreement between theory and simulations.

\section{Theory of Ostwald ripening with coalescence}
\label{sor}

In this Section we present a new theory of the globally-conserved (interface-controlled) Ostwald
ripening that accounts for coalescence. One of our assumptions is that each domain can be
represented by an equivalent \textit{circular} domain, or droplet, the area of which is equal to
the area of the domain. We shall denote by $f(R,t)$ the distribution function of the droplets with
respect to their radii. $f(R,t)$ is normalized by the condition $\int_{0}^{\infty}f(R,t)dR=n(t)$,
where $n(t)$ is the number density of the droplets. We start with a brief review of the
``classical" theory that neglects coalescence and goes back to Wagner \cite{Wagner}. Then we
derive a kinetic equation that accounts of coalescence. We shall focus on the long-time,
self-similar asymptotic solutions to that kinetic equation, find the solution by an iteration
procedure and compare it with the result of the phase-field simulations.

\subsection{Ostwald ripening without coalescence: a brief review}
At a late stage of coarsening $|H(t)|\ll 1$, so there is no nucleation of new domains. Then,
neglecting coalescence, one can write a simple continuity equation in $R$-space for the size
distribution function of domains, or droplets:
\begin{equation}\label{C1}
  \partial_t f +\partial_R\, (\dot{R}f)=0\ .
\end{equation}
When criterion (\ref{newth15}) is satisfied, the dynamics are describable by the area-preserving
motion by curvature (\ref{newth11}) (where we put $\delta=1$). This leads immediately to
\begin{equation}\label{C2}
\dot{R}= \frac{1}{R_c(t)}-\frac{1}{R}\ ,
\end{equation}
where the time-dependent critical radius $R_c(t)=\sqrt{2}/(3\,|H(t)|)$ is determined, at a late
stage of coarsening, by the conservation of the total area of the minority phase:
\begin{equation}\label{C3}
\pi \int_0^{\infty}\, R^2f\, dR=\varepsilon=const\,.
\end{equation}
Equations (\ref{C1})-(\ref{C3}) represent the classical model of interface-controlled Ostwald
ripening. This model was formulated by Wagner \cite {Wagner} by analogy with the theory of
Lifshitz and Slyozov \cite{LS} developed for the locally conserved (diffusion-controlled) dynamics.
Using Eqs. (\ref{C1})-(\ref{C3}), one obtains
\begin{equation}\label{C4}
R_c(t)=\frac{\int_0^{\infty}\, Rf\, dR}{\int_0^{\infty}\, f\, dR}= \langle R(t)\rangle \,,
\end{equation}
where $\langle R(t) \rangle$ is the time-dependent average radius of the droplets.

Droplets with $R>R_{c}(t)$ grow at the expense of droplets with $R<R_{c}(t)$ which shrink. The
late-time asymptotic behavior described by Eqs. (\ref{C1})-(\ref{C3}) is the following
\cite{Wagner}.  The critical radius grows with time (this corresponds to the decrease with time of
the effective magnetic field which plays the role of supersaturation). As a result, a droplet
which was growing at an early time begins to shrink at a later time. Since all the quantities are
position-independent, this model represents a mean-field theory. It should be noticed that the
mean-field approximation is much more accurate for the globally-conserved (interface-controlled)
Ostwald ripening than for the locally-conserved (diffusion-controlled) Ostwald ripening \cite{MS}.
First, in the globally-conserved case, the "mean field" $H(t)$ is the actual field in the system.
This is in contrast to the diffusion-controlled Ostwald ripening \cite {LS}, where a mean field
description of the supersaturation is an approximation valid only when the typical distances
between the droplets are very large compared to the typical droplet radius. The second difference
concerns the role of correlations. In the locally-conserved case, correlations between droplets
result from the Laplacian screening effect, and their relative contribution to the size
distribution function is of order $\varepsilon^{1/2}$ (see, \textit{e.g.} Ref. \cite{Marder}). The
effect of coalescence scales like $\varepsilon$ (see below) so, at small $\varepsilon$,
correlation effects should be much less significant. By contrast, in the interface-controlled case
direct correlations between droplets are exponentially small, and significant correlations can be
caused only by coalescence events.  Therefore, in the interface-controlled case, it is legitimate
to account for coalescence while neglecting correlations.

Wagner \cite{Wagner} obtained a self-similar solution to Eqs. (\ref{C1})-(\ref{C3}) (the Wagner
distribution)  that corresponds to a long-time asymptotics of the initial-value problem. The
similarity Ansatz is
\begin{equation}\label{C4a}
f(R,t)=\frac{1}{t^{3/2}}\Psi_{\beta}\left(\frac{R}{t^{1/2}}\right),
\;\;\;R_c(t)=\frac{t^{1/2}}{\beta}\,,
\end{equation}
where $\beta$ is a constant number. The scaled distribution $\Psi_{\beta}(\xi)$ obeys an ordinary
differential equation:
\begin{equation}\label{C4b}
\left(-\frac{\xi}{2}+\beta-\frac{1}{\xi}\right)\Psi_{\beta}^\prime(\xi)+
\left(-\frac{3}{2}+\frac{1}{\xi^2}\right)\Psi_{\beta}(\xi)=0 \,.
\end{equation}
The total area conservation (\ref{C3}) leads to normalization condition
\begin{equation}\label{C4c}
\pi \int_0^{\infty}\, \xi^2\, \Psi_{\beta}(\xi)\, d\xi  =\varepsilon = const\,.
\end{equation}
Formally solving Eq. (\ref{C4b}), one actually obtains a \textit{family} of solutions
parameterized by $\beta$. For $\sqrt{2}\leq \beta\leq 2\sqrt{2/3}$ these solutions have compact
support: they are positive on an interval $0<\xi<\xi_{max} (\beta)$, and zero elsewhere. Similar
solutions in 3D were investigated in Refs. \cite{MS,AMS,GMS}. We call these solutions localized.
For $0<\beta<\sqrt{2}$ the solutions of Eq. (\ref{C4b}) are extended: they have an infinite tail.
These solutions can be written as $const \cdot\Psi_{0\beta} (\xi)$, where
$$\Psi_{0\beta}(\xi)= \frac{\xi}{(\xi^2-2\beta\,\xi+2)^2}\times $$
\begin{equation}
\exp\left( -\frac{2\beta}{\sqrt{2-\beta^2}}\arctan\frac{\xi-\beta}{\sqrt{2-\beta^2}}\right)\,.
\label{C11(5)}
\end{equation}
Extended solutions fall off like $\xi^{-3}$ as $\xi \to \infty$. As a result, the integral in Eq.
(\ref{C4c}) diverges logarithmically, so the extended solutions are non-normalizable. Still, as we
shall see, they play a crucial role in the theory of Ostwald ripening with coalescence.

Which of the similarity solutions is selected by the dynamics (that is, represents a long-time
asymptotics of the initial value problem)?  It turns out that selection is ``weak", that is,
determined by the initial conditions. The Wagner distribution is selected for (normalizable)
\textit{extended} initial distributions. On the contrary, if the initial distribution $f(R,t=0)$
has compact support, one of the \textit{localized} distributions is selected. The selection is
determined by the asymptotics of $f(R,t=0)$ near the upper edge of its support \cite{MS,AMS,GMS}.

However, this weak selection rule was obtained in the framework of the classical formulation of
the problem, Eqs. (\ref{C1})-(\ref{C3}). One can expect that \textit{strong} selection (independent
of the initial conditions) can be obtained if one goes beyond the classical formulation. Indeed,
it was shown in Ref. \cite{Meerson} (see also \cite{zaltzman}) that an account of fluctuations
leads to strong selection. Fluctuations produce a tail in the time-dependent distribution function
and drive the solution towards the Wagner distribution. We shall see in the following that an
account of coalescence also leads to strong selection, even in the absence of fluctuations.

\subsection{Kinetic equation with coalescence}
We shall now take into account the processes of binary
coalescence. Coalescence events occur when two droplets contact
each other. Within the framework of the GLE, the positions of the
droplet centers remain fixed. Therefore, for coalescence to
happen, at least one of the droplets must be expanding. Consider a
droplet of radius $R_1<R<R_1+\Delta R_1$. The number density of
such droplets is $f(R_1,t)\Delta R_1$. Now consider another
droplet of radius $R_2<R<R_2+\Delta R_2$ in the vicinity of the
first droplet. If $\dot{R}_1+\dot{R}_2>0$ then, during the time
interval $\Delta t$, the distance between the boundaries of these
droplets will decrease by $(\dot{R}_1+\dot{R}_2)\Delta t$.  If the
distance $r$ between the centers of the droplets obeys the double
inequality
\begin{equation}
(R_1+R_2)\, \leq r\leq \,(R_1+R_2)+(\dot{R}_1+\dot{R}_2)\Delta t
\,,
\label{ring}
\end{equation}
(which assumes that the condition $\dot{R}_1+\dot{R}_2>0$ is
fulfilled), then these two droplets will collide during the time
interval $\Delta t$. Therefore, for the two droplets to collide,
the center of the second droplet should be located within a
circular ring, concentric with the first droplet, with radius
$R_{1}+R_{2}$ and width $(\dot{R}_{1}+\dot{R}_{2})\Delta t$. The
area of this ring is equal to
\begin{equation}\label{C5a}
2\pi\, (R_1+R_2)\, (\dot{R}_1+\dot{R}_2)\Delta t \ .
\end{equation}
Hence, the average number of such second droplets is equal to $$2\, M(R_1,R_2) f(R_2,t)\Delta
R_2\Delta t\ ,$$ where
$$M(R_1,R_2)=\pi\, (R_1+R_2)\, (\dot{R}_1+\dot{R}_2)\,
\theta(\dot{R}_1+\dot{R}_2)\ . $$ The total number of the collision events per unit area is equal
to
\begin{equation}\label{C5}
[2\, M(R_1,R_2)\, f(R_2,t)\Delta R_2\Delta t]\, f(R_1,t)\Delta R_1\, .
\end{equation}
Each collision leads to coalescence: disappearance of a droplet of radius $R_1$ and a droplet of
radius $R_2$, and creation of a new droplet.

Now we make two assumptions that will enable us to construct a
closed theory. First, we assume that the area of a new droplet,
formed by a binary coalescence event, is equal to the sum of the
areas of the two merging droplets. Second, we assume that new
droplet \textit{instantaneously} becomes circular \cite{Circular},
so its radius is $(R_1^2+R_2^2)^{1/2}$. The kinetic equation for
the size distribution function includes the rates of gains and
losses of droplets by coalescence. This leads to the following
equation:
$$\partial_t f +\partial_R\, (\dot{R}f)  =
-\frac{1}{2}\int\limits_0^\infty\,\int\limits_0^\infty\,\{2M(R_1,R_2)\times $$
$$
\left[\delta(R-R_1)+\delta(R-R_2)-\delta\left(R-\sqrt{R_1^2+R_2^2}\;\right)\right]\times
$$
\begin{equation}\label{C6}
f(R_1,t)\, f(R_2,t)\, dR_1\, dR_2 \} \,,
\end{equation}
where $\delta\,(...)$ is the Dirac's delta-function and the factor $1/2$ is introduced in order to
avoid counting each coalescence event twice. Performing integration with $\delta (R-R_1)$ and
$\delta (R-R_2)$ and taking into account the symmetry of $M(R_1,R_2)$ under a transposition of its
arguments:
$$M(R_1,R_2)=M(R_2,R_1)\,,$$ we obtain:
 $$\partial_t f+\partial_R\,
(\dot{R}f)  =$$ $$ = -2 f(R,t) \int\limits_0^\infty \, M(R,R_1)
f(R_1,t)\, dR_1 + $$ $$ +\int\limits_0^\infty\int\limits_0^\infty\
\,  M(R_1,R_2)\, \delta\left(R-\sqrt{R_1^2+R_2^2}\;\right)\times$$
\begin{equation}\label{C7}
 f(R_1,t)\, f(R_2,t)\, dR_1\, dR_2 \,.
\end{equation}
Integration of the right hand side of Eq.~(\ref{C7}) over $R^2\, dR$ yields zero, so the new
kinetic equation preserves the conservation law (\ref{C3}) as it should. In addition, the simple
relation (\ref{C4}) continues to hold. Integrating the right hand side of Eq.~(\ref{C7}) over
$dR$, and over $R\, dR$, respectively, one can show that the coalescence term reduces the number
density of the droplets and the total interface length. Moreover, the new equation preserves the
dynamic scaling. Indeed, if some $f(R,t)$ and $R_c(t)$ give a solution to Eqs.~(\ref{C7}),
(\ref{C3}) and (\ref{C4}), then $f^{\prime}(R,t)=\eta^3f(\eta R,\eta^2t)$ and $R_c^{\prime}
(t)=\eta^{-1} R_c(\eta^2 t)$ give another solution to the same equations. This invariance under a
stretching transformation implies the existence of a self-similar solution that will be considered
in the next subsection.

While deriving Eq. ~(\ref{C7}), we neglected effects of interactions of three droplets. By this we
refer to cases where there are three closely lying droplets. In these cases triple coalescence
events may occur. In addition, an excluded area in the ring (\ref{ring}) appears. The effects of
this excluded area, and of the triple coalescence events were not taken into account in our
theory. These effects are expected to be of order $\varepsilon^{3}$, while the effects of binary
coalescence events are of order $\varepsilon^{2}$. Therefore, Eq.~(\ref{C7}) is expected to be
valid for small area fractions $\varepsilon$. We shall see, however, that a very good accuracy is
obtained even for the moderate value of $\varepsilon=0.25$ used in our simulations, when triple
coalescence events do occur (see Fig. \ref{fig2}).

Another limitation of our theory concerns the large-$R$ tail of $f(R,t)$. The tail shape is
affected by higher-order coalescence events unaccounted for in our theory. This limitation is not
very important in practice. The main contribution to the critical radius $R_c$ comes, for
normalizable distributions, by the ``body" of the distribution function, rather than by the tail.

We conclude this subsection by a brief discussion of a different type of coalescence: Brownian
coalescence. Following the pioneering work of Smoluchowski \cite{Smoluchowski}, Binder and Stauffer
\cite{Binder} suggested a mean-field scenario of phase separation in alloys in which clusters of
the minority phase are regarded as Brownian particles: they perform random walk in space. When two
clusters collide, they merge into a larger single cluster. The corresponding kinetic equation
includes an integral term whose general structure resembles that of the integral term in Eq.
(\ref{C7}), but with a different kernel $M(R_1,R_2)$. If the cluster diffusivity is a power-law
function of the cluster size, one arrives at a self-similar solution for the size distribution
function of the droplets. An important further development was the work of Siggia \cite{Siggia}
who considered hydrodynamic interactions between randomly moving and coalescing droplets in phase
separating binary fluids. Following the work of Siggia, the Brownian coalescence in binary fluids
has been extensively studied theoretically and experimentally. Among important issues here is a
crossover from Ostwald ripening (the Lifshitz-Slyozov-Wagner mechanism) to Brownian coalescence
\cite{Bray94,Siggia,Haas}, plethora of hydrodynamic interactions in the process of coalescence
\cite{Siggia,Tanaka,Nikolayev}, scaling violations \cite{Yeomans} etc. In parallel,  Brownian
coalescence has been investigated in the context of coarsening of clusters of atoms or vacancies
diffusing on surfaces, following particle deposition \cite{Kandel}. It is clear that Brownian
coalescence is different in its nature from the coalescence process considered in this work. In
contrast to Brownian coalescence, droplets in our system do not move: they coalesce only because
they grow.

\subsection{Self-similar solution with coalescence}

Equations (\ref{C7}) and (\ref{C3}) admit the same similarity Ansatz as Eqs.~(\ref{C1}) and
(\ref{C3}):
\begin{equation}\label{C8}
f(R,t)=\frac{1}{t^{3/2}}\Phi\left(\frac{R}{t^{1/2}}\right)
\end{equation}
and
\begin{equation}\label{C9}
  R_c(t)=\beta^{-1}\, t^{1/2} 
\,,
\end{equation}
where $\beta$ is again an unknown yet constant number. The scaled
distribution function $\Phi(\xi)$ obeys the following
integro-differential equation: $$
\left(-\frac{\xi}{2}+\beta-\frac{1}{\xi}\right)\Phi^\prime(\xi)+
\left(-\frac{3}{2}+\frac{1}{\xi^2}\right)\Phi(\xi)= $$ $$ -2
\Phi(\xi)\, \int\limits_0^\infty\,w(\xi,\xi_1)\, \Phi(\xi_1)\,
d\xi_1+ $$
\begin{equation}\label{C10}
  \int\limits_0^\infty\int\limits_0^\infty\, w(\xi_1,\xi_2)\, \delta\left(\xi-\sqrt{\xi_1^2+\xi_2^2}\,\right)\,
  \Phi(\xi_1)\Phi(\xi_2)\, d\xi_1\, d\xi_2
\end{equation}
subject to normalization condition
\begin{equation}\label{C11}
\pi \int_0^{\infty}\, \xi^2\, \Phi(\xi)\, d\xi  =\varepsilon \,.
\end{equation}
In Eq. (\ref{C10}) we denoted
$$
w(\xi_1,\xi_2)=\pi\, (\xi_1+\xi_2)\, \left(2\beta-\frac{1}{\xi_1}- \frac{1}{\xi_2}\right)\, \times
$$
\begin{equation}
\theta\left(2\beta-\frac{1}{\xi_1}- \frac{1}{\xi_2}\right)\, , \label{Add1}
\end{equation}
where $\theta\,(...)$ is the theta-function. It is convenient to rewrite Eq.~(\ref{C10}) in a
symbolic form:
\begin{equation}
{\cal L}_\beta\Phi = {\cal N}_\beta [\Phi] \,, \label{C11(1)}
\end{equation}
where
\begin{equation}
{\cal L}_\beta\Phi (\xi)= \left(-\frac{\xi}{2}+\beta-\frac{1}{\xi}\right)\Phi^\prime(\xi)+
\left(-\frac{3}{2}+\frac{1}{\xi^2}\right)\Phi(\xi) \ , \label{C11(1_1)}
\end{equation}
and $$ {\cal N}_\beta\,\left[\Phi\,\right](\xi)=-2 \Phi(\xi)\,
\int\limits_0^\infty\,w(\xi,\xi_1)\, \Phi(\xi_1)\, d\xi_1+ $$
\begin{equation}
  \int\limits_0^\infty\int\limits_0^\infty\,
  w(\xi_1,\xi_2)\, \delta\left(\xi-\sqrt{\xi_1^2+\xi_2^2}\,\right)\,
 \Phi(\xi_1)\Phi(\xi_2)\, d\xi_1\, d\xi_2\, .
\label{C11(1_2)}
\end{equation}

One important property of Eq. (\ref{C10}) can be noticed immediately: the coalescence term
vanishes identically at $0\leq\xi<1/(2 \beta)$. As a result, the scaled distribution function at
$0\leq\xi<1/(2 \beta)$ should coincide (up to a $\xi$-independent multiplier) with one of the
solutions of the classical Wagner's problem. A simple argument shows that parameter $\beta$,
parameterizing this solution, should be less than $\sqrt{2}$. Indeed, inverting the linear
operator ${\cal L}_\beta$, we rewrite Eq. (\ref{C11(1)}) as an integral (rather than
integro-differential) equation:
\begin{equation}
\Phi (\xi)=\Psi_{\beta}(\xi)\, \left[\int\limits_{\xi}^{\infty}\, \frac{ {\cal N}_{\beta}\,
\left[\Phi(\xi^{\prime})\right]\,
d\xi^{\prime}}{\left(\frac{\xi^\prime}{2}-\beta+\frac{1}{\xi^\prime}\right)
\Psi_{\beta}(\xi^{\prime})} + C_{1}\right]\,, \label{inteq}
\end{equation}
where functions $\Psi_{\beta}(\xi)$ were introduced in subsection A and $C_{1}$ is a constant.
Unless $\beta<\sqrt{2}$, the integral over $d\xi^{\prime}$ diverges. Therefore, $\Psi_{\beta}(\xi)$
should be one of the \textit{extended} solutions $\Psi_{0\beta}(\xi)$, given by Eq.
(\ref{C11(5)}). In addition, since the second term in the square brackets of Eq. (\ref{inteq})
would lead to divergence of the integral appearing in the normalization condition (\ref{C11}), we
must choose $C_{1}=0$. Hence, Eq. (\ref{inteq}) reads:
\begin{equation}
\Phi (\xi) = \Psi_{0\beta}(\xi)\, \int\limits_{\xi}^{\infty}\,
\frac{ {\cal N}_{\beta}\, \left[\Phi(\xi^{\prime})\right]\,
d\xi^{\prime}}{\left(\frac{\xi^\prime}{2}-\beta+\frac{1}{\xi^\prime}\right)
\Psi_{0\beta}(\xi^{\prime})} \,.
 \label{inteq1}
\end{equation}
Integral equation (\ref{inteq1}) and normalization condition (\ref{C11}) make a complete set. For
a given $\varepsilon$, the scaled distribution function $\Phi = \Phi_{\beta}(\xi)$ and parameter
$\beta =\beta(\varepsilon)$ are uniquely determined. Therefore, an account of coalescence does
provide strong selection to the problem of Ostwald ripening.

\subsection{Solving Eqs.~(\protect\ref{inteq1}) and (\protect\ref{C11})} \label{iteration}

Our procedure for solving Eqs.~(\ref{inteq1}) and (\ref{C11}) employs the one-to-one
correspondence between $\varepsilon$ and $\beta$. Therefore, one can fix $\beta$ and solve
Eq.~(\ref{inteq1}) by iterations for $\Phi=\Phi_{\beta}(\xi)$. The normalization condition
(\ref{C11}) is \textit{not} used at this stage. After a sufficiently accurate estimate for
$\Phi=\Phi_{\beta}(\xi)$ is obtained, one employs Eq. (\ref{C11}) to calculate $\varepsilon$ that
corresponds to this $\beta$. Repeating this procedure for different $\beta$, one obtains the
family of solutions $\Phi_{\beta}(\xi)$ and the dependence $\varepsilon=\varepsilon (\beta)$.
Inverting this dependence, one arrives at $\beta=\beta(\varepsilon)$ and finds the correspondence
between $\varepsilon$ and scaled distributions $\Phi_{\beta}(\xi)=\Phi_{\beta (\varepsilon)}(\xi)$.

Now we introduce an iteration scheme that implements this idea. The scheme exploits the fact that,
for the exact solution of Eq.~(\ref{inteq1}),
\begin{equation}
\Phi(\xi)=\chi \,\Psi_{0\beta}(\xi)\mbox{~~~~at~~~~}0\leq\xi<\frac{1}{2\beta}\, , \label{C11(9_3)}
\end{equation}
where $\chi=\chi(\beta)$ is $\xi$-independent and unknown in advance. Let us introduce an auxiliary
unknown function $\phi(\xi)= \chi^{-1} \Phi (\xi)$ \cite{phi}. By definition,
\begin{equation}\label{phi1}
\phi(\xi)= \Psi_{0\beta}(\xi)\mbox{~~~~at~~~~}0\leq\xi<\frac{1}{2\beta}\,.
\end{equation}
In terms of the new function $\phi$, Eq.~(\ref{inteq1}) becomes
\begin{equation}
\phi (\xi) = \chi \,\Psi_{0\beta}(\xi)\, \int\limits_{\xi}^{\infty}\, \frac{ {\cal N}_{\beta}\,
\left[\phi(\xi^{\prime})\right]\,
d\xi^{\prime}}{\left(\frac{\xi^\prime}{2}-\beta+\frac{1}{\xi^\prime}\right)
\Psi_{0\beta}(\xi^{\prime})} \,.
 \label{inteq2}
\end{equation}
The iteration scheme for Eq.~(\ref{inteq2}) is the following:
$$\phi_{k+1}(\xi)=$$
\begin{equation}
\chi_{k}\Psi_{0\beta}(\xi)\!\int\limits_{\xi}^\infty\! \frac{{\cal N}_\beta[\phi_{k}(\xi^\prime)]\,
d\xi^\prime}{
 \left(\frac{\xi^\prime}{2}-\beta+\frac{1}{\xi^\prime}\right)
\Psi_{0\beta}(\xi^\prime)}\,,  \label{C11(9_7)}
\end{equation}
for $k=0,1,2,\dots\,.$ We use the arbitrariness of $\chi_k$ and demand that, at each iteration, Eq.
(\ref{phi1}) is satisfied:
\begin{equation}
\phi_{k+1}(\xi)=\Psi_{0\beta}(\xi)\mbox{~~~~at~~~~}0\le\xi<\frac{1}{2\beta}\, . \label{C11(9_8)}
\end{equation}
This implies
\begin{equation}
\chi_k=\left[\int\limits_{1/(2 \beta)}^\infty\frac{{\cal N}_\beta[\phi_{k}(\xi)]\, d\xi}{
 \left(\frac{\xi}{2}-\beta+\frac{1}{\xi}\right)\,
\Psi_{0\beta}(\xi)} \right]^{-1}\,, \label{C11(9_9)}
\end{equation}
for $k=0,1,2,\dots\,.$ Equation~(\ref{C11(9_9)}) and  (\ref{C11(9_7)}) define the iteration process
explicitly. If the sequence $\phi_k\,, k=0,1,2,\dots$ converges to a finite limit $\phi(\xi)$,
then the sequence $\chi_k\,, k=0,1,2,\dots$ converges to a finite (positive) number $\chi$, and we
can find $\Phi(\xi)=\chi\,\phi(\xi)$. What is left is to use Eq. (\ref{C11}) and compute the
corresponding $\varepsilon$. If the convergence of the iteration scheme is fast enough, then
\begin{equation}
\Phi_{k}(\xi)=\chi_k \phi_k (\xi) \label{200}
\end{equation}
and
\begin{equation}
\varepsilon_k(\beta)=\pi\, \chi_k \int \xi^2 \phi_k (\xi)\, d \xi\,,\;\; k=1,2,\dots\, .
\label{epsk}
\end{equation}
give a good approximation to the solution already after a small number of iterations.

We implemented this iteration procedure numerically. As it is clear from Eqs.~(\ref{C11(9_7)}),
(\ref{C11(9_9)}), (\ref{200}) and (\ref{epsk}), the numerics involve only calculations of (double
and triple) definite integrals. We started with the trial function $\phi_0(\xi) =\Psi_{0\beta}
(\xi)$. The advantage of this trial function is that it already satisfies  Eq. (\ref{phi1}). We
performed a detailed investigation of the convergence of this scheme for different values of
$\beta$ up to iteration $k=8$. The results of this investigation (and proofs of convergence of the
integrals at infinity) will be presented in a separate publication. The main result is that, with
this choice of the trial function, the convergence is very fast in the body of the scaled
distribution function. For example, in the case of $\varepsilon=0.25$ the first iteration already
gives an accurate result. Convergence in the tail is a more subtle issue that will be addressed
separately.

The $\beta(\varepsilon)$-dependence, found by this procedure is shown in Fig. \ref{fig6}. One can
see that, as $\varepsilon \to 0$, $\beta \to \sqrt{2}$ from below. Overall, this behavior is
expected. Coalescence effects become small at small area fractions and, as $\varepsilon \to 0$,
the scaled distribution function should approach the Wagner distribution. It is surprising,
however, that for quite small $\varepsilon$ (for example, $0.01$), the value of $\beta$ is still
significantly different from $\sqrt{2}$.

Figure \ref{fig7} shows the scaled distribution function for $\beta=0.93$ (which corresponds to
$\varepsilon \simeq 0.25$). The solid line shows the result of the first iteration
$\Phi_{1}(\xi)$. This result agrees well with the phase field simulations, even at the moderately
large value of $\varepsilon=0.25$. This is a strong evidence in favor of the major role of
coalescence in the interface-controlled Ostwald ripening. The dashed line is the trial function
$\Phi_{0}(\xi)$. One can see that this trial function (with $\beta=0.93$) already gives a fairly
accurate estimate to the solution in the body of the scaled distribution. Therefore,  the
non-normalizable extended distributions (\ref{C11(5)}) do play a special role in the theory. In the
tail region the solution falls off more rapidly than $\Psi_{0\beta}$, so there is no problem with
normalization condition (\ref{C11}). The zero moment $M_0$ corresponding to the once-iterated
numerical solution (the solid line in Fig. \ref{fig7}) is equal to $5.08 \cdot 10^{-2}$ which is
in good agreement with the value $4.72 \cdot 10^{-2}$ obtained in the simulations.

An independent estimate of $\beta$ at $\varepsilon=0.25$ is provided by the dynamics of the
effective "magnetic field" $H(t)$. The area-preserving dynamics, Eq.~(\ref{newth11}), imply that
$|H(t)| \simeq |\sqrt{2}\,\bar{\kappa}/3|$. Assuming that all droplets have circular shape (an
assumption already used in our theory), we have $|\kappa|= 1/\langle R(t)\rangle$. Employing the
similarity solution, we arrive at $|\kappa|=\beta/t^{1/2}$. Therefore, $1/|H(t)|\sim c\,t^{1/2}$,
where $c=3/(\sqrt{2}\beta)$. For the corrected power-law fit shown in Fig.~\ref{fig5} $c=2.3$. This
corresponds to $\beta\simeq 0.92$ which is remarkably close to the value of $0.93$ obtained by the
iteration procedure.

\section{Conclusions}
\label{conclusions}

We investigated globally conserved phase ordering dynamics in systems with short range
correlations. The numerical simulations were done with a 2D phase-field model: a Ginzburg-Landau
equation with a global conservation law. The sharp-interface limit of this equation: the
area-preserving motion by curvature was introduced, and a criterion for its validity formulated.
Assuming dynamic scaling within the model of area-preserving motion by curvature, we obtained the
$1/2$ dynamic exponent for the characteristic coarsening length scale (for critical and
off-critical quench), and for the effective "magnetic field" (for off-critical quench). Our
numerical simulations for critical quench and for an off-critical quench with area fraction
$\varepsilon=0.25$ confirm these scaling laws. The results for critical quench coincide with
earlier results, obtained by particle simulations. The scaled form of the equal-time
pair-correlation function is found to weakly depend on the area fraction, similar to the
locally-conserved systems.

Recently, dynamic scale invariance and ``normal" scaling have been reported in the same globally
conserved coarsening system, but for \textit{long-range} (power-law) correlations in the initial
conditions \cite{PCM}. Our present results, combined with those of Ref. \cite{PCM}, indicate
``normal" scaling properties of globally conserved systems for any generic initial conditions.

The main focus of this work was on the dynamics of the size distribution function of the minority
phase domains for the off-critical quench. Our phase field simulations show that this distribution
function exhibits, at large times, a self-similar form. The scaled distribution function, however,
is dramatically different from the well-known Wagner distribution, despite the fact that
correlations between domains are negligible in globally-conserved systems. We attributed this
strong deviation to coalescence that provides, at small $\varepsilon$, a leading correction to the
Wagner's theory. We developed a new theory of Ostwald ripening that takes into account binary
coalescence events. The theory possesses dynamic scale invariance and yields a non-linear
integro-differential equation for the scaled distribution function. For a given area fraction, the
problem has a unique solution. Therefore, coalescence renders a strong selection rule to the
problem of Ostwald ripening: the scaled distribution is selected by the area fraction of the
minority domains. The scaled distribution function predicted by the new theory is in good
agreement with the scaled distribution obtained from the phase field simulations for a moderate
area fraction of 0.25. In addition, the new theory accurately predicts the amplitude of the
late-time power law of the effective magnetic field $H(t)$. Deviations from the classical Wagner's
theory remain significant even for very low area fractions. Therefore, coalescence plays a major
role in the interface-controlled Ostwald ripening.

We hope that the results of this work will stimulate further experiments on globally-conserved
interface-controlled systems.

\section*{Acknowledgments}

We are grateful to B.P. Vollmayr-Lee for useful discussions. This work was supported in part by the
Israel Science Foundation, administered by the Israel Academy of Sciences and Humanities, and by
the Russian Foundation for Basic Research (grant No. 99-01-00123).

\section*{Appendix. Criteria for the area-preserving dynamics}

Let us find the general conditions under which the area-preserving dynamics [Eq. (\ref{newth11})]
represent an accurate approximation to the more general sharp-interface theory [Eqs. (\ref{newth6})
and (\ref{newth7})]. We first notice that in the general sharp-interface theory it is the quantity
$A(t)-L^2 H(t)/4$, rather than $A(t)$, that is conserved \cite{MS}. Hence, a necessary condition
for the validity of the area-preserving dynamics is simply
\begin{equation}
H(t)\ll \frac{A(t)}{L^{2}}\,. \label{criterion1}
\end{equation}
Additional criteria are found in the following manner. We notice that Eq. (\ref{newth7}) includes
the same combination $\delta\overline{k}+(\delta\mu)^{1/2} gH(t)$ as the one that appears on the
right hand side of Eq. (\ref{newth9}). Therefore a constancy or, more precisely, slow variation of
$A(t)$ implies a slow variation of $H(t)$. Correspondingly, the term on the left hand side of Eq.
(\ref{newth7}) should be small in this case compared to each of the two terms on the right hand
side. We can exploit this fact and formally solve Eq. (\ref{newth7}) perturbatively:
$H(t)=H^{(0)}(t)+h(t)$, where $H^{(0)}(t)=-g^{-1}(\delta/\mu)^{1/2}\;\overline{\kappa}$, is the
leading term, and $h(t)$ is the subleading term. Substituting $H(t)$ into Eq. (\ref{newth7}) and
keeping terms up to order $h(t)$ we obtain:
\begin{equation}
h(t)=\frac{L^{2}}{4g\,{(\delta\mu)}^{1/2}\,\Lambda(t)}\dot{H}^{(0)}(t)=-\frac{
L^{2}}{4g^{2}\mu\,\Lambda(t)}\dot{\overline{\kappa(s,t)}}
 \,.
\label{newth12}
\end{equation}
The perturbation expansion is valid as long as
$$
|h(t)|\ll |H^{(0)}(t)|\,\,\,\,\,\, \mbox{and} \,\,\,\,\,\,\frac{|\dot{h(t)}|}{(\delta
\mu)^{1/2}\,\Lambda(t)} \ll \frac{|h(t)|}{L^2}\,.
$$
Using the expressions for $H^{(0)}(t)$ and $h(t)$ we see that the dynamics described by (\ref
{newth6}) and (\ref {newth7}) are (approximately) area-preserving if the following two inequalities
hold:
\begin{equation}
\frac{|\;\dot{\overline{\kappa(s,t)}}\;|}{(\delta\mu)^{1/2}\,\Lambda(t)}\ll
\frac{\;|\overline{\kappa(s,t)}|\;}{L^{2}}
 \,,
\label{newth13}
\end{equation}
and
\begin{equation}
\left|\frac{d}{dt}\left[\;\frac{\dot{\overline{k(s,t)}}\;}
{(\delta\mu)^{1/2}\,\Lambda(t)}\right]\right|\ll \frac{|\;\dot{\overline{k(s,t)}}\;|}{L^{2}}
 \,.
\label{newth14}
\end{equation}
Now we can see why the area-preserving dynamics do not apply to a single circular droplet. In this
case the zero order of the perturbation theory would give
$\overline{\kappa(s,t)}=1/R=\mbox{const}$, where $R$ is the droplet radius. Therefore, criterion
(\ref{newth14}) is violated.

For a ``donut" one has $\overline{\kappa(s,t)}=0$. Therefore, neither of the two criteria
(\ref{newth13}) and (\ref{newth14}) is obeyed, and the dynamics are not area-preserving. If the
system consists of $N_{1} (t)$ domains of the majority phase and $N_2$ (t) domains of the minority
phase, the Gauss-Bonet theorem \cite{diffgeom} yields
\begin{equation}
\overline{\kappa(s,t)}=\frac{2\pi(N_{1}(t)-N_{2}(t))}{\Lambda(t)} \,, \label{eqtopo}
\end{equation}
where $N_{1}(t)$ does not include the large "sea" of the majority phase. We shall employ this
relation in the following.

Let us check criteria (\ref{criterion1}), (\ref{newth13}) and (\ref{newth14}) for the ``standard
problem" of phase-ordering, when the initial conditions describe a disordered state with
short-ranged correlations. We assume that the system exhibits, at late times, dynamic scale
invariance. In other words, the characteristic domain size $l(t)$ is the only relevant length
scale. For an off-critical quench, the phase-ordering morphology is that of Ostwald ripening:
competition of droplets of the minority phase, see below. This implies the following scaling
relations: $\overline{\kappa(s,t)}\sim 1/l(t)$, $\dot{\overline{k(s,t)}}\sim\delta/l^{3}(t)$, and
$\Lambda(t)\sim N_{2}(t)l(t)\sim \varepsilon L^{2}/l(t)$. In addition, $H(t)\sim \lambda/l(t)$ and
$A=\varepsilon\, L^2$. Using these relations in any of the inequalities (\ref{criterion1}),
(\ref{newth13}) or (\ref{newth14}), we arrive at
\begin{equation}
\frac{\lambda}{l(t)}\ll \varepsilon
 \,.
\label{newth15}
\end{equation}
That is, the more general sharp-interface theory is reducible to the area-preserving motion by
curvature as long as the ratio between the interface width and the coarsening length scale is much
smaller than the area fraction of the minority phase. If the coarsening system remains a two-phase
system, and if the system size is big enough, this condition holds, at late times, for any nonzero
area fraction \cite{stripe}. Notice that, at $\varepsilon \ll 1$ it takes more time for the system
to reach the area-preserving regime.

\pagebreak

\begin{figure}
\hspace{-0.1in} \center{\epsfysize=4.5in \epsffile{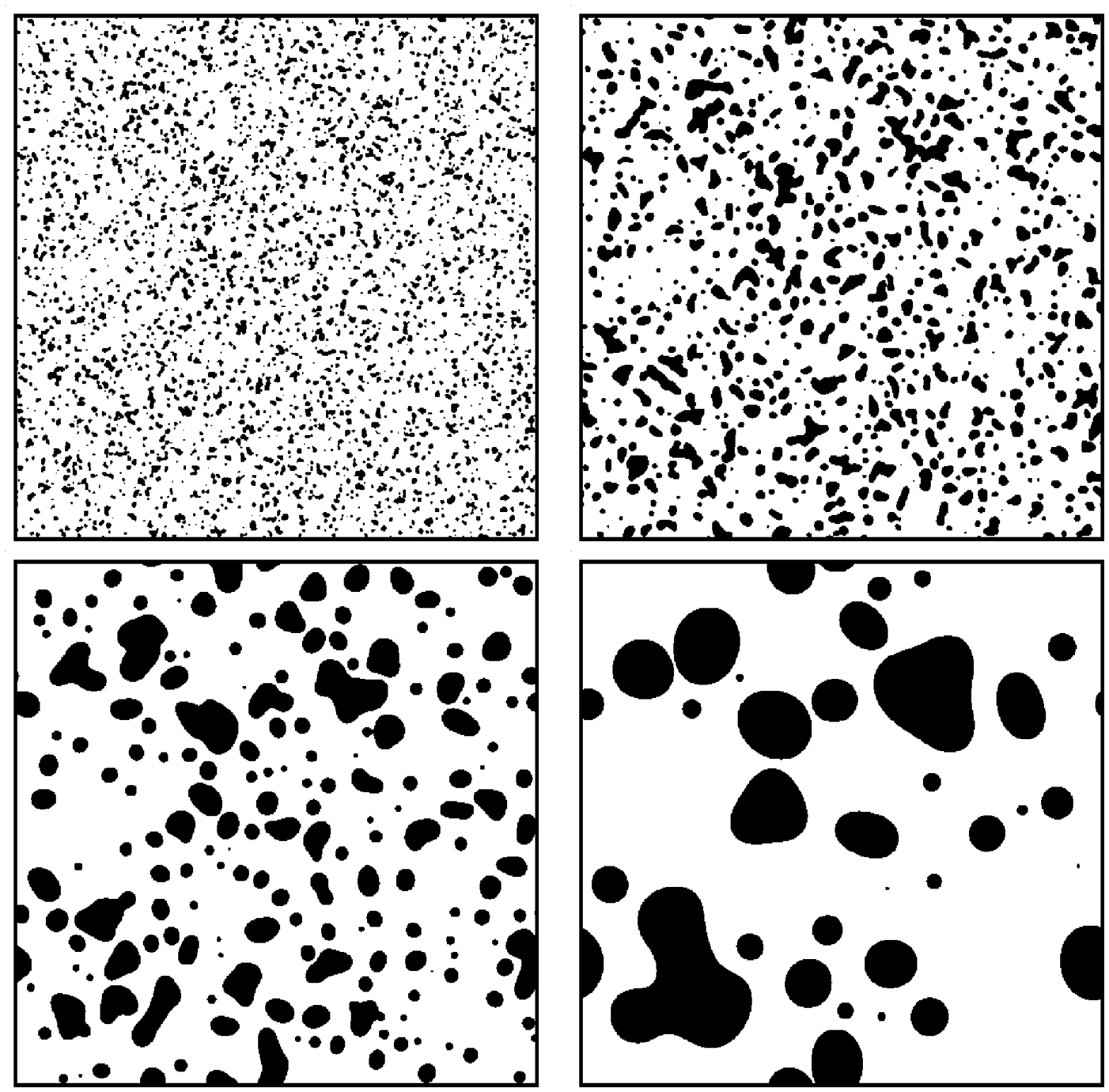}}\vspace{-1.3in}\caption{Snapshots
of globally conserved  coarsening for off-critical quench with area fraction $\varepsilon=0.25$.
The upper row corresponds to $t=5.2$ (left) and 32.3 (right), the lower row to $t=204.8$ (left)
and 1305.5 (right).}\label{fig2}
\end{figure}

\begin{figure}
\hspace{-0.1in} \center{\epsfysize=4.3in \epsffile{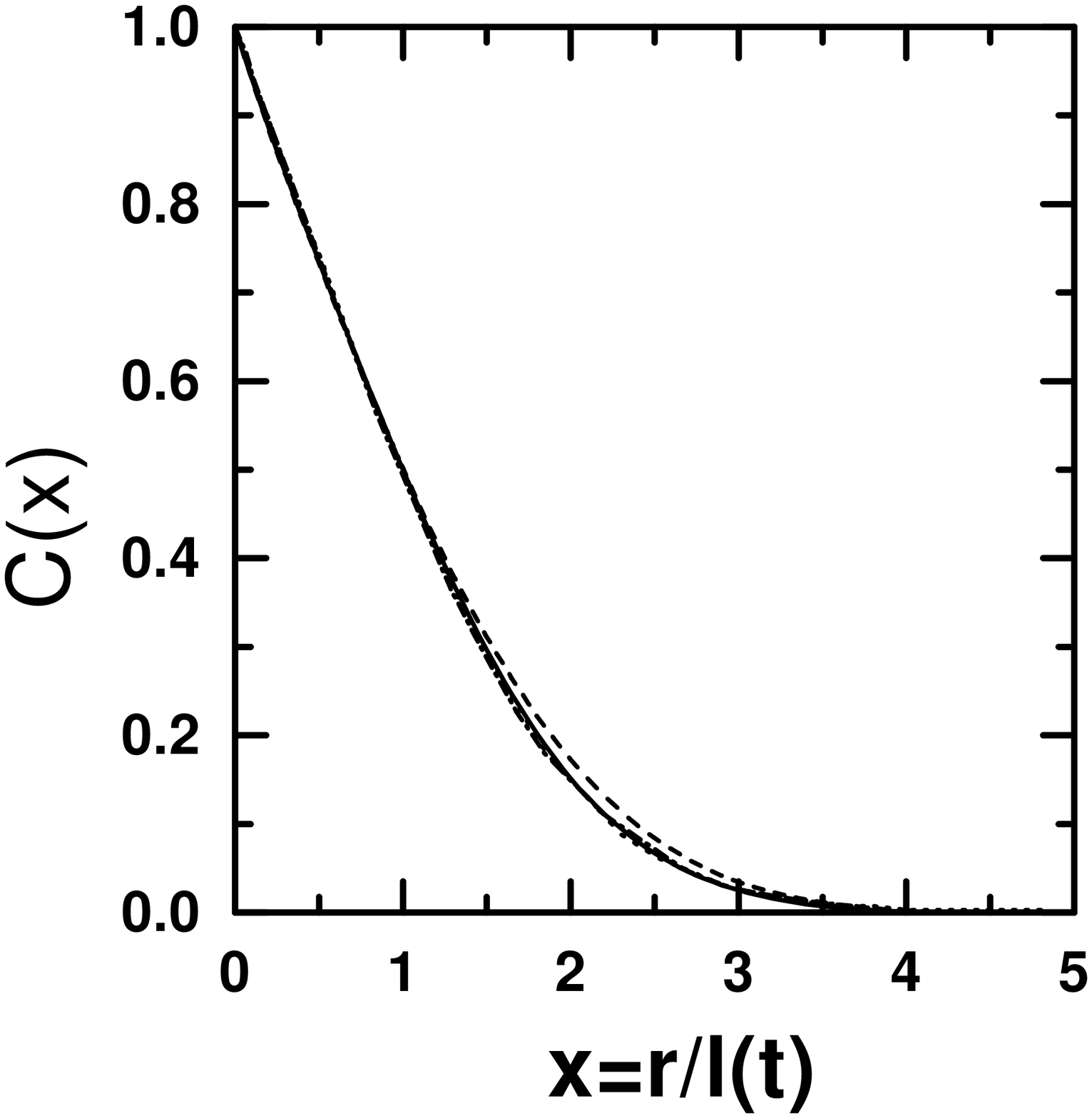}}\vspace{-1.1in} \caption{Scaled
correlation function obtained by numerical simulations with the GLE subject to a global
conservation law for times $t>15$. The solid line is $C(x)$ for critical quench, and the dashed
line is $C(x)$ for off-critical quench with area fraction $\varepsilon=0.25$. The dotted
(dashed-dotted) line represents $C(x)$ obtained in particle simulations of globally conserved
\protect\cite{Rutenberg} (non-conserved \protect\cite{Humayun}) dynamics for critical
quench.}\label{fig3}
\end{figure}

\begin{figure}
\hspace{-0.1in} \center{\epsfysize=4.3in
\epsffile{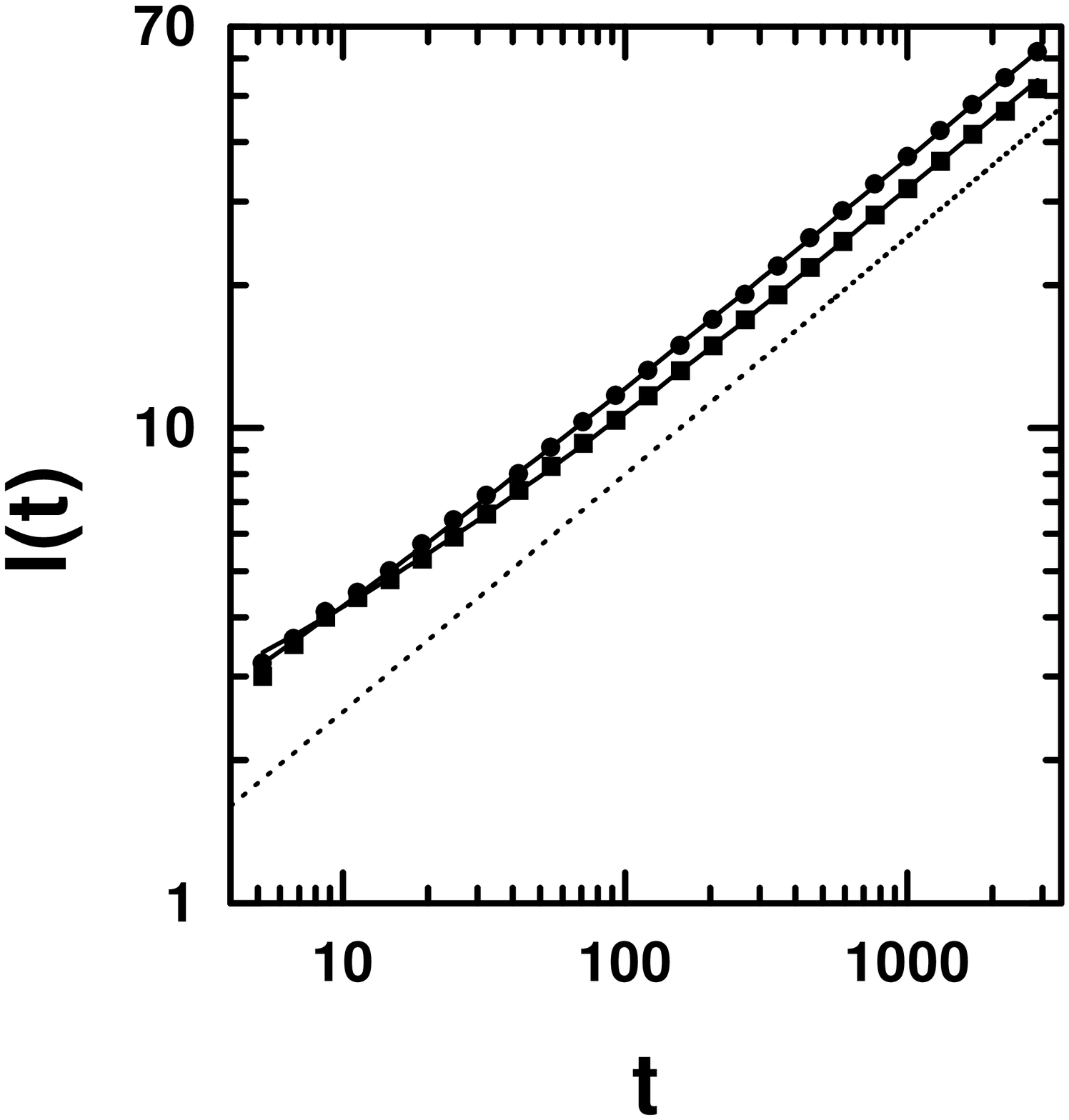}}\vspace{-1.2in}\caption{Characteristic coarsening length $l$ vs. time for
the critical quench (circles), and off-critical quench with area fraction $\varepsilon=0.25$
(squares). The two solid lines are corrected power-law fits $l(t)=l_{0}+bt^{\alpha}$ with
$\alpha=0.50$, $l_{0}=0.5$, and $b=1.2$ for the critical quench, and $\alpha=0.51$, $l_{0}=1.3$,
and $b=0.9$ for the off-critical quench. The dotted line shows a pure $t^{1/2}$ power
law.}\label{fig4}
\end{figure}

\begin{figure}
\hspace{-0.1in} \center{\epsfysize=4.3in
\epsffile{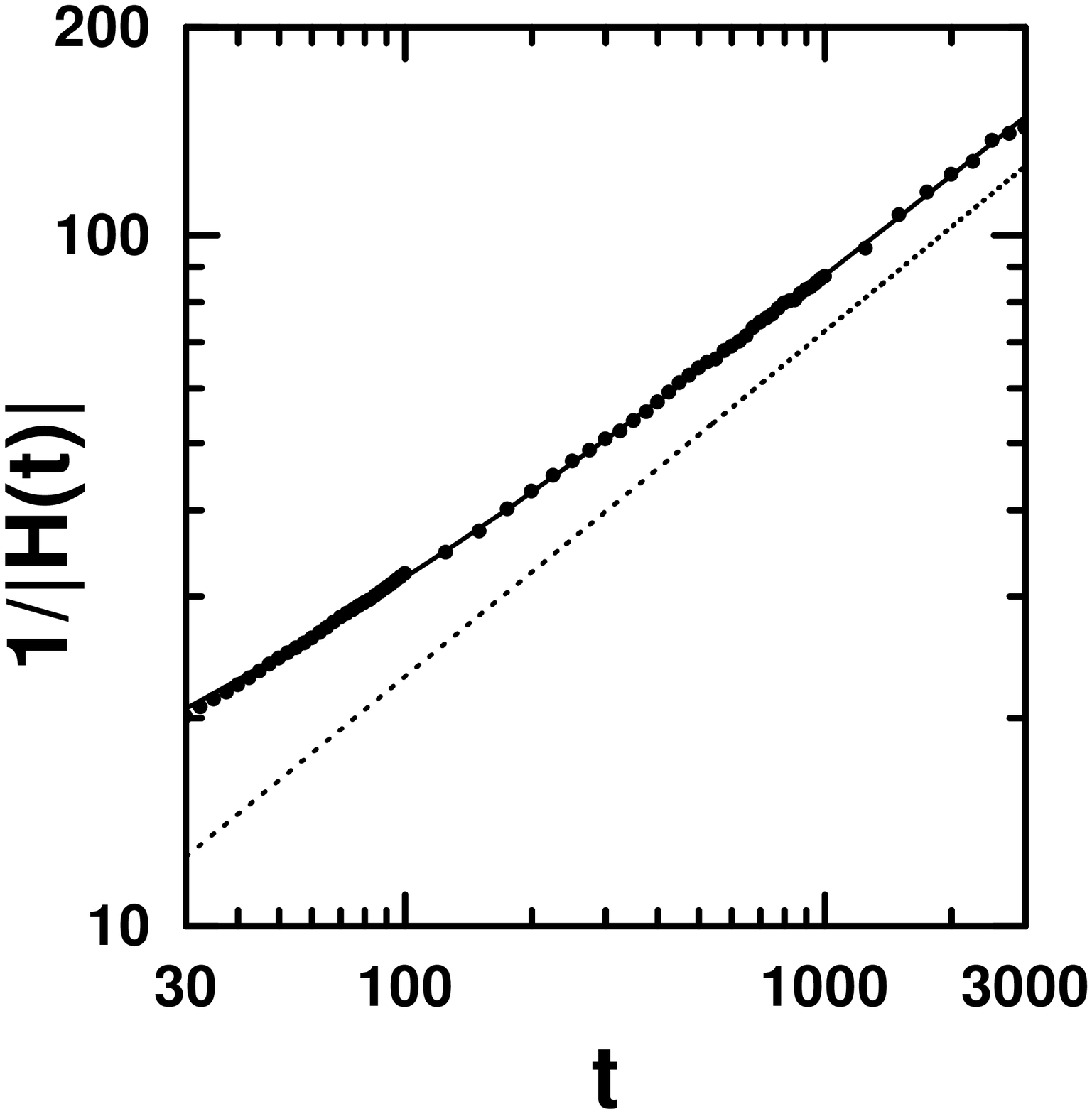}}\vspace{-1.3in}\caption{$1/|H(t)|$ vs. time for an off-critical quench
with area fraction $\varepsilon=0.25$ (circles). The solid line is a corrected power law fit
$1/|H(t)|=a+ct^{\alpha}$ with $a=7.4$, $c=2.3$, and $\alpha=0.51$. The dotted line is $2.3\,
t^{1/2}$ power law.} \label{fig5}
\end{figure}

\begin{figure}
\hspace{-0.1in}\center{\epsfxsize=3.0in \epsfysize=4.5in
\epsffile{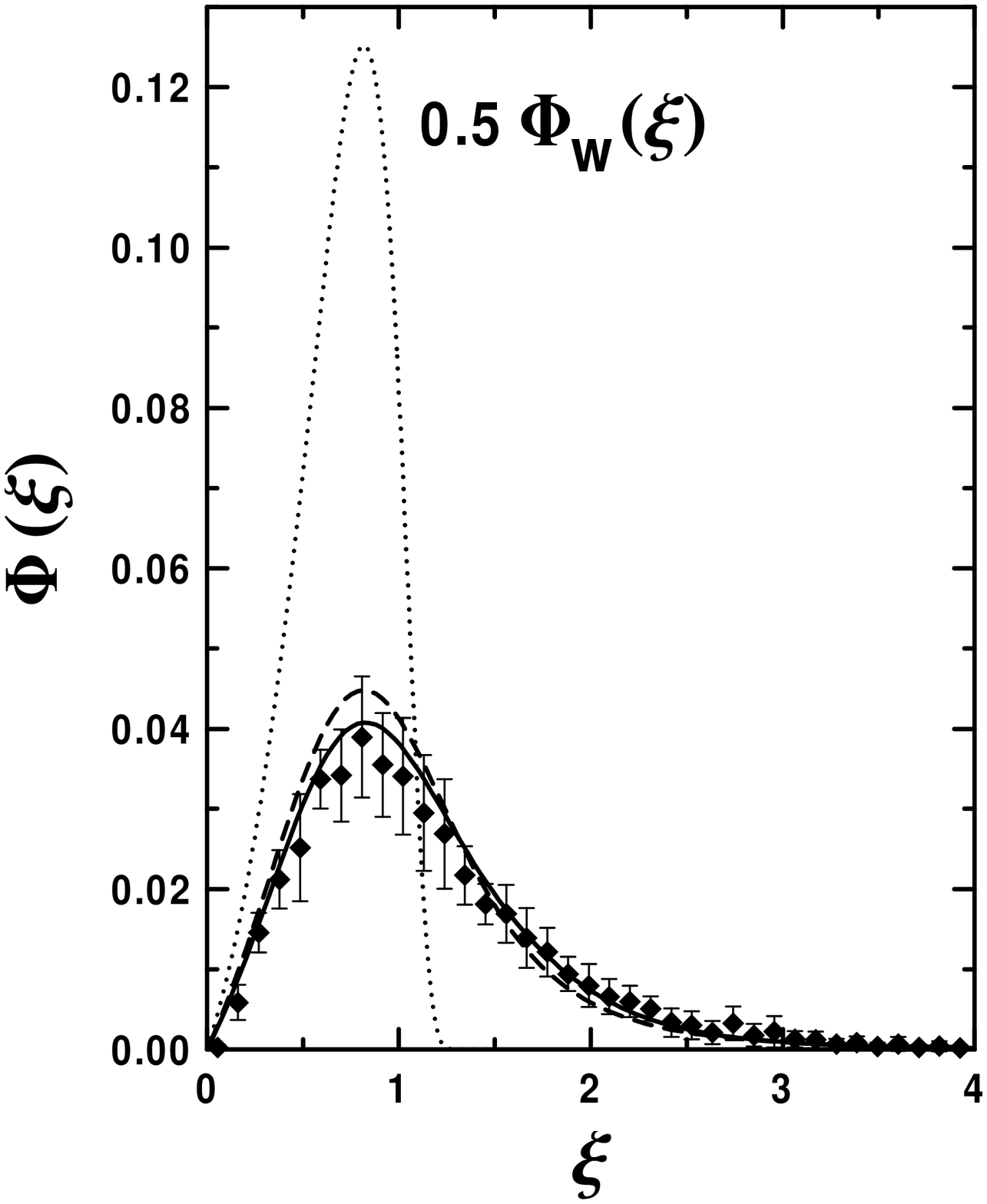}}\vspace{0.0in}\caption{Scaled distribution function of domain sizes
$\Phi(\xi)$, where $\xi=R/t^{1/2}$. The diamonds represent $\Phi_{num}(\xi)$: the scaled
distribution obtained in the simulations with the GLE for times $t>120$. The error bars show
(twice) the variance of the scaled distribution functions for 13 time moments in the interval
$120<t<2900$. The dotted line is the Wagner distribution $\Phi_W (\xi)$, Eq.
(\protect\ref{wagner}), for the same area fraction $\varepsilon=0.25$. In order to show it on the
same graph with $\Phi_{num}$ we had to multiply it by $0.5$. The dashed and solid lines show the
distributions $\Phi_{0}(\xi)$ and $\Phi_{1}(\xi)$, respectively, predicted by the theory of
Ostwald ripening with coalescence, presented in Sec. \protect\ref{sor}. These distributions
represent the zero and first iterations of the iteration procedure [see Sec. \protect\ref{sor},
Eq. (\ref{200})] for $\beta=0.93$ which corresponds to $\varepsilon \simeq 0.25$.} \label{fig7}
\end{figure}

\begin{figure}
\hspace{-0.2in} \center{\epsfxsize=3.0in \epsfysize=2.7in \epsffile{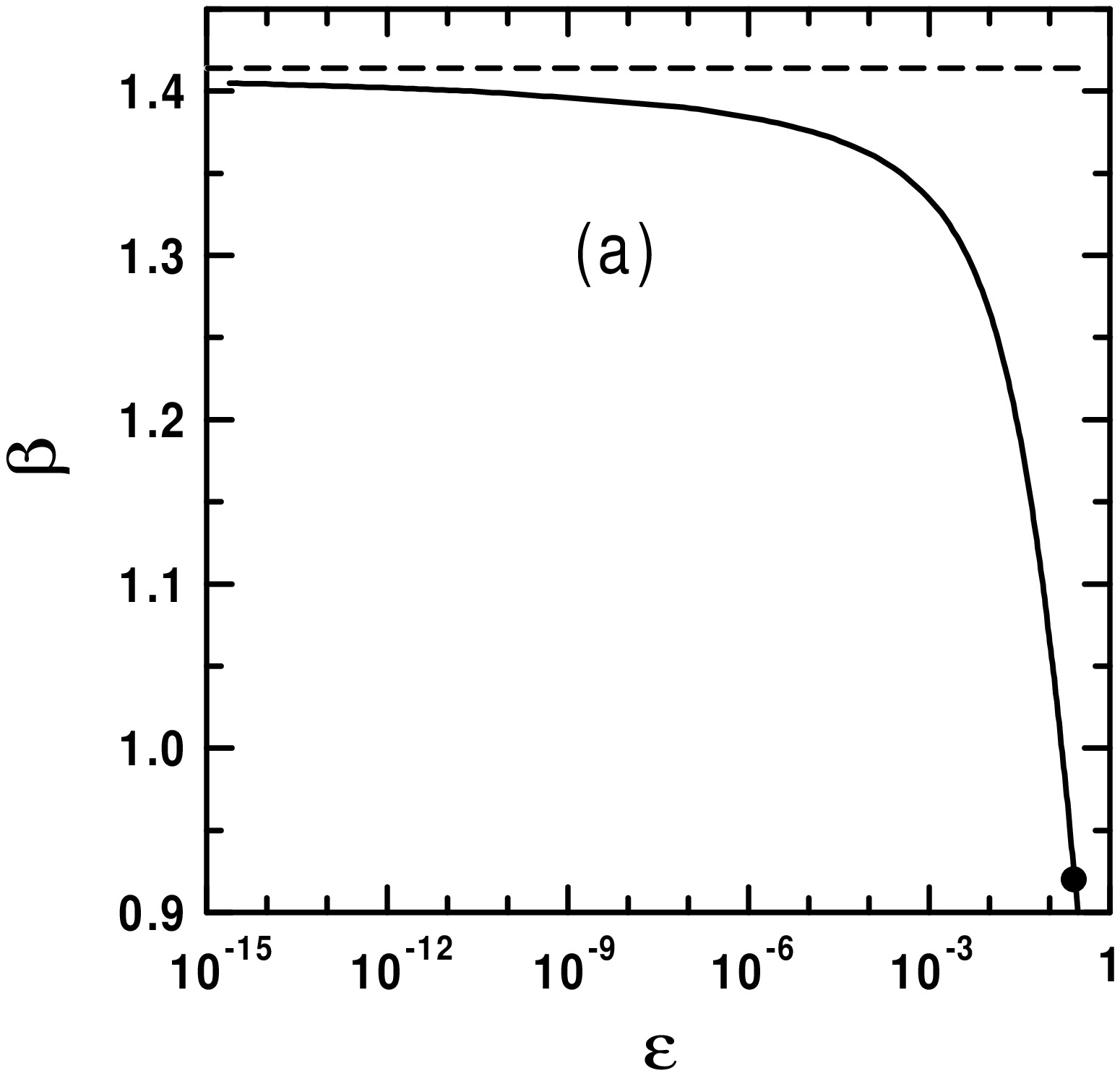}} \hspace{-0.1in}
\vspace{0.3in}\center{\epsfxsize=3.0in \epsfysize=2.7in \epsffile{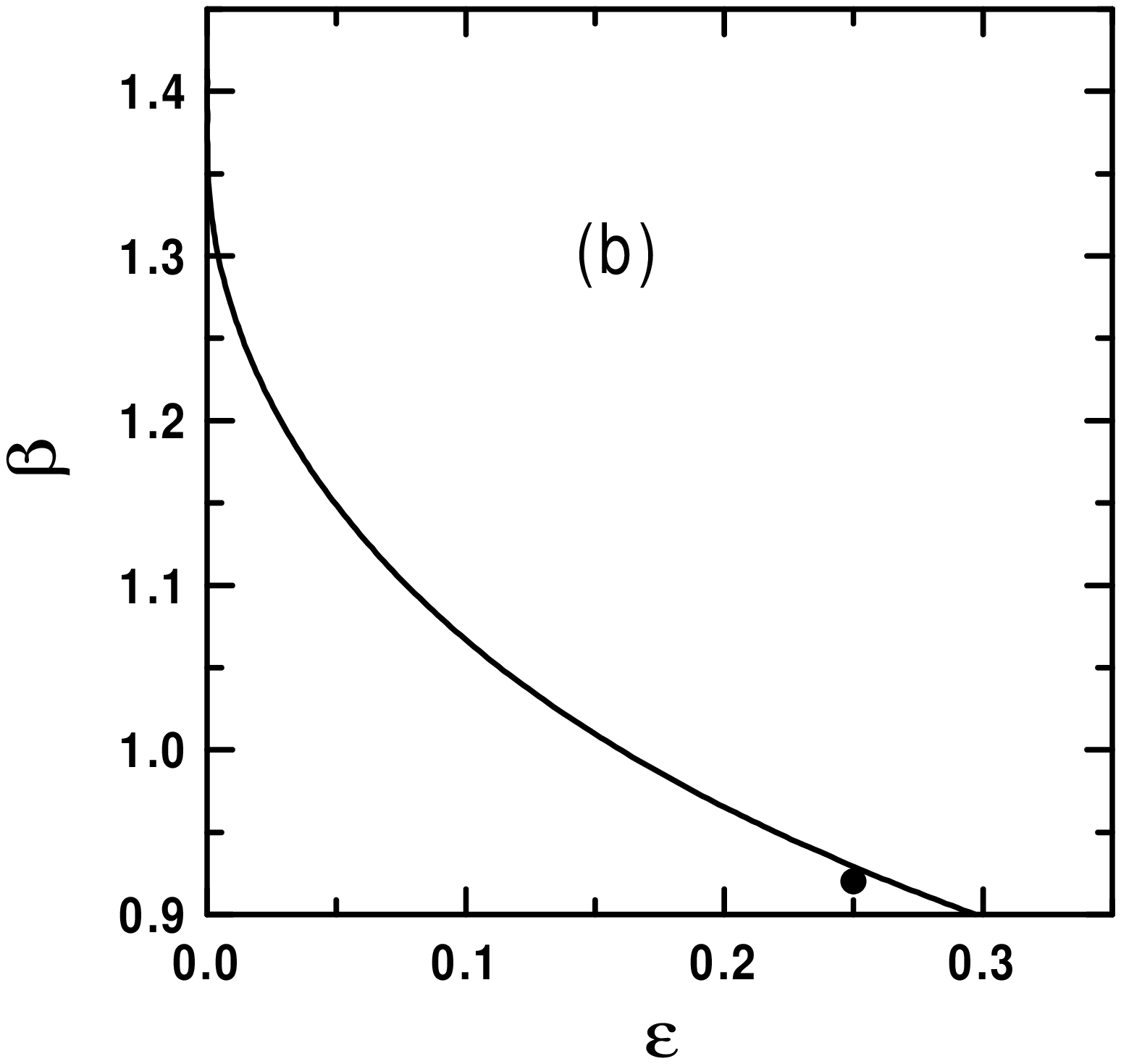}}
\vspace{0.0in}\caption{Parameter $\beta$ versus $\varepsilon$ (solid lines) as predicted by the
theory of Ostwald ripening with coalescence. This dependence was obtained by a single iteration
applied to Eqs.~(\protect{\ref{inteq1}}) and (\protect{\ref{C11}}). The dashed line shows the
limiting value $\beta=\sqrt{2}$ expected at $\varepsilon \to 0$. The solid circle is the point
found in our phase-field simulations. Figure (a) shows $\varepsilon$ in a logarithmic scale,
figure (b) in a linear scale.} \label{fig6}
\end{figure}


\begin{references}

\bibitem{Bray94} A.J. Bray,  Adv. Phys. {\bf 43}, 357 (1994).

\bibitem{Schimansky} L. Schimansky-Geier, C. Z\"{u}licke and E. Sch\"{o}ll, Z. Phys. B-Cond.
Mat. \textbf{84}, 433 (1991).

\bibitem{Rubinstein} J. Rubinstein and P. Sternberg, IMA J. Appl.
Math. {\bf 48}, 249 (1992).
\bibitem{Majumdar} C. Sire and S.N. Majumdar, Phys. Rev. Lett.
{\bf 74}, 4321 (1995); Phys. Rev. E {\bf 52}, 244 (1995).
\bibitem{MS} B. Meerson and P.V. Sasorov, Phys. Rev. E {\bf
53}, 3491 (1996).
\bibitem{Rutenberg} A.D. Rutenberg, Phys. Rev. E {\bf 54}, 972 (1996).
\bibitem{PCM} A. Peleg, M. Conti, and B. Meerson, Phys. Rev. E \textbf{64}, 036127 (2001).
\bibitem{Wynblatt} P. Wynblatt and N.A. Gjostein, in: {\it
Progress in Solid State Chemistry}, edited by J.O. McCaldin and G. Somojrai, (Pergamon, Oxford,
1975) Vol. 9, p. 21.
\bibitem{Zinke-Allmang} M. Zinke-Allmang, L.C. Feldman, and
M.H. Grabow, Surface Sci. Reports {\bf 16}, 277 (1992).
\bibitem{Williams} Z. Toroczkai and E. Williams,
Phys. Today {\bf 52} (12), 24 (1999).
\bibitem{Aranson1} I. S. Aranson, D. Blair D, V.A. Kalatsky, G.W. Crabtree, W.K. Kwok,
V.M. Vinokur, and U. Welp, Phys. Rev. Lett. {\bf 84}, 3306 (2000).
\bibitem{AMSV} I. Aranson, B. Meerson, P.V. Sasorov and V.M. Vinokur, e-print cond-mat/0107443.
\bibitem{Bray91} A.J. Bray, Phys. Rev. Lett. {\bf 66}, 2048
(1991), and references therein.
\bibitem{simulations} J.F. Annett and J.R. Banavar, Phys. Rev. Lett.
{\bf 68}, 2941 (1992); L.L. Moseley, P.W. Gibbs, and N. Jan, J.
Stat. Phys. {\bf 67}, 813 ( 1992).
\bibitem{Lee} B.P. Lee and A.D. Rutenberg, Phys. Rev. Lett. {\bf
79}, 4842 (1997).
\bibitem{Glazier} J.A. Glazier, M.P. Anderson, and G.S. Grest,
Philos. Mag. B {\bf 62}, 615 (1990).
\bibitem{Grest} G.S. Grest, M.P. Anderson, and D.J. Srolovitz,
Phys. Rev. B {\bf 38}, 4752 (1988).
\bibitem{LS} I.M. Lifshitz and V.V. Slyozov, Zh. Eksp. Teor. Fiz. \textbf{35}, 479 (1958) [Sov. Phys.
JETP \textbf{8}, 331 (1959)]; J. Phys. Chem. Solids, {\bf 19}, 35 (1961).
\bibitem{Wagner} C. Wagner, Z. Elektrochem. {\bf 65}, 581 (1961).
\bibitem{AMS} I. Aranson, B. Meerson, and P.V. Sasorov, Phys. Rev. E {\bf
52}, 948 (1995).
\bibitem{Marder}  M. Marder, Phys. Rev. A {\bf 36}, 858 (1987).
\bibitem{Mikhailov} A.S. Mikhailov, {\it Foundations of Synergetics
I. Distributed Active Systems} (Springer-Verlag, Berlin, 1993).
\bibitem{PMVC} A. Peleg, B. Meerson, A. Vilenkin, and M. Conti,
Phys. Rev. E {\bf63}, 066101 (2001).
\bibitem{Gage} M. Gage, Contemp. Math. {\bf 51}, 51 (1986).
\bibitem{Allen-Cahn} S.M. Allen and J.W. Cahn, Acta Metall. {\bf
27}, 1085 (1979).
\bibitem{MM} B. Meerson and I. Mitkov, Phys. Rev. E {\bf
54}, 4644 (1996).
\bibitem{Ostwald} W. Ostwald, Z. Phys. Chem. \textbf{34}, 495 (1900).
\bibitem{Rogers} T.M. Rogers and R.C. Desai, Phys. Rev. B \textbf{39}, 11956 (1989).
\bibitem{Fratzl} P. Fratzl, J.L. Lebowitz, O. Penrose, and J. Amar, Phys. Rev.
B \textbf{44}, 4794 (1991).
\bibitem{Humayun} K. Humayun and A.J. Bray, J. Phys. A {\bf 24},
1915 (1991); Phys. Rev. B {\bf 46} 10594 (1992).
\bibitem{Elder} K.R. Elder and R.C. Desai, Phys. Rev. B \textbf{40}, 243 (1989).
\bibitem{mathematica} S. Wolfram, {\it The Mathematica Book}, 3rd ed.
(Cambridge Univ. Press, 1996), p. 1081.
\bibitem{GMS} B. Giron, B. Meerson, and P.V. Sasorov, Phys. Rev. E
{\bf 58}, 4213 (1998).
\bibitem{Meerson} B. Meerson, Phys. Rev. E \textbf{60}, 3072 (1999).
\bibitem{zaltzman} I. Rubinstein and B. Zaltzman, Phys. Rev. E \textbf{61}, 709 (2000).
\bibitem{Circular} In fact, new droplets just formed by coalescence
do not have a circular form (see Fig. \ref{fig2}). However, for small $\varepsilon$, deviations
from the circular form are significant for big droplets, that is in the tail of the size
distribution function. In the ``body" of the size distribution function (which gives the dominant
contribution to the critical radius $R_c$), the droplets are already close to circular.
\bibitem{Smoluchowski} M. von Smoluchowski, Z. Phys. Chem. \textbf{92}, 129 (1917).
\bibitem{Binder} K. Binder and D. Stauffer, Phys. Rev. Lett. \textbf{33}, 1006 (1974).
\bibitem{Siggia} E.D. Siggia, Phys. Rev. A \textbf{20}, 595 (1979).
\bibitem{Haas} C.K. Haas, J.M. Torkelson, Phys. Rev. E \textbf{55}, 3191 (1997).
\bibitem{Tanaka} H. Tanaka, J. Chem. Phys. \textbf{105}, 10099 (1996).
\bibitem{Nikolayev} V.S. Nikolaev, D. Beysens and P. Guenoun, Phys. Rev. Lett. \textbf{76}, 3144 (1996).
\bibitem{Yeomans} A.J. Wagner and J.M. Yeomans, Phys. Rev. Lett. \textbf{80}, 1429 (1998).
\bibitem{Kandel} C.R. Stoldt, C.J. Jenks, P.A. Thiel, A.M. Cadilhe and J.W. Evans, J.
Chem. Phys. \textbf{111}, 5157 (1999);  D. Kandel, Phys. Rev. Lett. \textbf{79}, 4238 (1997).
\bibitem{phi} One can check that a straightforward iteration scheme for $\Phi(\xi)$
would diverge.
\bibitem{diffgeom} E. Kreyszig, {\it Differential Geometry} (University of
Toronto Press, Toronto, 1959).
\bibitem{stripe} A criterion similar to Eq. (\ref{newth15}) also appears when the initial condition
represents a single long stripe \cite{PMVC}. In that case $l(t)$ is the stripe length.

\end{references}
\end{document}